\documentclass[useAMS,usenatbib]{mn2e}
\usepackage{epsfig}
\usepackage{float}
\usepackage{placeins}
\usepackage{graphicx}
\usepackage{epsfig}
\usepackage{bm}
\usepackage{amsfonts}
\usepackage{amssymb}
\usepackage{times}
\usepackage{natbib}
\usepackage{color}

\title[Flux and spectral variations in S5 0716+714]{Multi-band optical variability of the Blazar S5 0716+714 in outburst state during 2014-2015}
\author[Agarwal et al.]
{Aditi Agarwal$^{1,2}$\thanks{E-mail: aditi@aries.res.in},
Alok C.\ Gupta$^{1,2}$\thanks{E-mail: acgupta30@gmail.com},
R. Bachev$^{3}$, 
A. Strigachev$^{3}$,
E. Semkov$^{3}$,
\newauthor Paul J. Wiita$^{4}$,
J. H. Fan$^{5}$,
U. S. Pandey$^{2}$,
S. Boeva$^{3}$,
B. Spassov$^{3}$
\\
\\ 
$^{1}$Aryabhatta Research Institute of Observational Sciences (ARIES),
Manora Peak, Nainital -- 263002, India\\
$^{2}$Department of Physics, DDU Gorakhpur University, Gorakhpur - 273009, India \\
$^{3}$Institute of Astronomy and National Astronomical Observatory,
Bulgarian Academy of Sciences, 72 Tsarigradsko Shosse Blvd., 1784 Sofia, Bulgaria \\
$^{4}$Department of Physics, The College of New Jersey, 2000 Pennington Rd., Ewing, NJ 08628-0718, USA\\
$^{5}$Center for Astrophysics, Guangzhou University, Guangzhou 510006, China \\
}
\begin{document}
\newdimen\digitwidth
\setbox0=\hbox{2}
\digitwidth=\wd0
\catcode `#=\active
\def#{\kern\digitwidth}

\date{Accepted ....... Received  ......; in original form ......}

\pagerange{\pageref{firstpage}--\pageref{lastpage}} \pubyear{2014}

\maketitle

\label{firstpage}

\begin{abstract}
We analyzed the multi-band optical behaviour of the BL Lacertae object, S5 0716+714, during its outburst state from 2014 November -- 2015
March. We took data on 23 nights at three observatories, one in India and two in Bulgaria, making quasi-simultaneous 
observations in B, V, R, and I bands. We  measured multi-band optical fluxes, colour and spectral variations for this blazar on intraday and short timescales.
The source was in a flaring state during the period analyzed and displayed intense variability in all wavelengths.
R band magnitude of 11.6 was attained by the target on 18 Jan 2015,
which is the brightest value ever recorded for S5 0716+714. 
The discrete correlation function method  yielded good correlation between the bands with
no measurable time lags, implying that radiation in these bands originate from the same region and by the same mechanism. 
We  also used the
structure function technique to look for characteristic timescales in the light curves.
During the times of rapid variability, no evidence for the source to display spectral changes with magnitude was found on either of the timescales.
The amplitude of variations tends to increase with increasing frequency with a maximum of $\sim$ 22\% seen during flaring states in B band.
 A mild trend of larger variability amplitude as the source brightens was also found.
 We found the duty cycle of our source during the analyzed period to be $\sim$ 90\%.
We also investigated the optical spectral energy distribution of  S5 0716+714 using B, V, R, and I data points for 21 nights. We briefly discuss
physical mechanisms most likely responsible for its flux and spectral variations.
 \end{abstract}
 
\begin{keywords}
galaxies: active --- BL Lacertae objects: general ---  quasars: individual -- BL Lacertae objects: individual: S5 0716+714
\end{keywords}

\section{Introduction}
\label{sec:Introduction}

Blazars constitute an important subset of radio-loud Active Galactic Nuclei (RLAGNs), consisting of both BL Lacertae objects (BL Lacs),
with largely featureless optical spectra, and flat spectrum radio quasars (FSRQs), which have prominent emission lines.
Blazars are marked by strong optical linear polarization ($>$3\%), 
and variability at 
wavelengths spanning the entire electromagnetic (EM) spectrum, from radio to gamma-ray bands. 
They are associated with Doppler boosted relativistic jets making an angle of $\lesssim$ 10$^{\circ}$ with the line of sight (LOS) 
(e.g.\ Urry \& Padovani 1995). 

\begin{table*}

\caption{ Details of telescopes and instruments}

\textwidth=6.0in

\textheight=9.0in

\vspace*{0.2in}

\noindent

\begin{tabular}{p{2.2cm}p{2.2cm}p{2.3cm}p{2.2cm}} \hline

Site:              &A                                 &B                                 & C           \\\hline

Telescope:       &1.30m Ritchey-Chr\'etien          &  60-cm Cassegrain                &50/70-cm Schmidt     \\

CCD model:       &Andor 2K CCD                      &  FLI PL09000                     &FLI PL160803       \\

Chip size:       &$2048\times2048$ pixels           & $3056\times3056$ pixels          & $4096\times4096$ pixels       \\

Pixel size:      &  $13.5\times13.5$ $\mu$m         & $12\times12$ $\mu$m              & $9\times9$ $\mu$m            \\

Scale:           &       0.535\arcsec/pixel         & 0.330\arcsec/pixel$^{\rm a}$     & 1.079\arcsec/pixel                   \\

Field:           & $18\arcmin\times18\arcmin$       &  $16.8\arcmin\times16.8\arcmin$   & $73.66\arcmin \times 73.66 \arcmin$    \\

Gain:            & 1.4 $e^-$/ADU                    & 1.0 $e^-$/ADU                     & 1.0 $e^-$/ADU            \\

Read Out Noise:  & 4.1 $e^-$ rms                    & 8.5 $e^-$ rms                     & 9.0 $e^-$ rms        \\

Binning used:    &    $2\times2$                    &  $3\times3$                       & $1\times1$                    \\
  
Typical seeing : &   1.2\arcsec to 3.0\arcsec       & 1.5\arcsec to 4.0\arcsec          & 2\arcsec to 4\arcsec           \\\hline

\end{tabular} \\

A  : 1.3-m Ritchey-Chretien Cassegrain optical telescope, ARIES, Nainital, India \\

B  : 60-cm Cassegrain telescope at Astronomical Observatory Belogradchik, Bulgaria \\

C  : 50/70-cm Schmidt telescope at National Astronomical Observatory, Rozhen, Bulgaria   \\

$^{\rm a}$ With a binning factor of $1\times1$

\end{table*}

Variability timescales in blazars range from a few minutes through days and months to decades (e.g.\ Carini \& Miller 1992).
Blazar variability often is arbitrarily divided into three classes: flux variations up to a few tenths 
of a magnitude over a timescale of few minutes to less than a day are called intra-day variability (IDV; e.g.\ Wagner \& Witzel 1995), 
or intra-night variability or microvariability; flux variations typically exceeding $\sim$1 mag on timescale of days to few months 
are known as short time variability (STV); 
while the changes from several months to many years with amplitude changes upto $\sim$5 mag are usually called long term variability 
(LTV; e.g.\ Gupta et al.\ 2004).
Measurements of variability amplitudes and duty cycles, temporal lags between bands, along with spectral changes, can provide information about the 
location, size, structure and dynamics of the regions emitting non-thermal  photons (e.g.\ Ciprini et al.\ 2003).

Blazars' broadband spectral energy distributions (SEDs) consist of two well-defined components (e.g.\ Mukherjee et al.\ 1997; 
Weekes 2003) extending from radio up to gamma-rays. 
The low energy spectrum (radio to UV/X-rays) is generally agreed to be synchrotron emission from the relativistic 
electrons of the jet (Ulrich, Maraschi, \& Urry 1997).
Inverse Compton scattering of the synchrotron emission itself and/or
external photons is normally held to be responsible for the high energy spectrum 
(e.g.\ Sikora \& Madejski 2001; B{\"o}ttcher 2002). The location of the first peak of the SED has been 
used to sub-classify blazars into low energy peaked blazars (LBLs) and high energy peaked blazars (HBLs). In case of LBLs 
the first hump peaks in the NIR/optical and the second component usually peaks at GeV energies, while in case of HBLs the first 
peak lies in UV/X-ray band and the second one at TeV  energies (e.g., Padovani \& Giommi 1995; Abdo et al.\ 2010). 
However, if the SED peaks are located at intermediate frequencies, it gives rise to the intermediate peaked blazar (IBL) classification (e.g.
Sambruna, Maraschi, \& Urry 1996).

The BL Lacertae object S5 0716+714 ($\alpha_{2000.0}$ = 07h 21m 53.4s, $\delta_{2000.0}$ = $+71^{\circ} 20^{'} 36.4^{\prime \prime}$), 
at redshift of 0.31$\pm$0.08 (Nilsson et al.\ 2008) is one of the most active blazars in optical bands, displaying flux variability on 
timescales from hours to days (e.g.\ Heidt \& Wagner 1996; Montagni et al.\ 2006;  
Gupta et al.\ 2008; and references therein). 
This BL Lacertae object has been classified as an IBL by Giommi et al.\ (1998) since the frequency of the first SED peak varies 
between 10$^{14}$ -- 10$^{15}$ Hz and the frequency of the high energy peak is near 1 GeV (Ferrero et al.\ 2006; 
Massaro et al.\ 2008). The concave X-ray spectrum between 0.1 and 10 keV band provides a signature of the presence of both the tail from 
the synchrotron emission and a flatter part from the inverse Compton (IC) spectrum, thus supporting its IBL nature (Ferrero 
et al.\ 2006). 

Wagner \& Witzel (1995) found the duty cycle of the source to be $\sim$ 1 implying that the source is always in active state. 
Along with its high declination and brightness, this means that  S5 0716+714 has been a target of a large number of intranight 
monitoring observations (e.g., Wu et al.\ 2005; 
Ostorero et al.\ 2006; Stalin et al.\ 2009; Gupta et al.\ 2008).
A high polarization of $\sim$20\% was found during the first optical polarization studies of this source by
Takalo, Sillanpaeae, \& Nilsson (1994) along with intra-day polarization variability of $\sim$3.5\%. Fan et al.\ (1997) reported an even higher
optical polarization of 29\%.  
Violent polarimetric variability in the source was reported by Ikejiri et al.\ (2011). Based on long term data and detection 
of optical outbursts, Gupta et al. (2008) reported a possible period of long term variability of $\sim$ 3.0 $\pm$ 0.3 years. 
S5 0716$+$714 has occasionally shown quasi-periodic oscillations (QPOs) in its time series data. Quirrenbach et al.\ (1991) 
reported a good evidence of a QPO of $\sim$ 1 day simultaneously in optical and radio bands. Gupta, Srivastava, \& Wiita
(2009) used wavelet analysis to study the excellent intraday optical LCs of the target obtained by Montagni et al.\ (2006) and 
found evidences of QPOs ranging between $\sim$ 25 \& $\sim$ 73 minutes on five different nights. Rani et al.\ (2010)
reported a QPO of $\sim$ 15 min in optical LCs of the source.
An upper limit of 
$\sim$ 10 min time lag between the variations in B and I bands was derived by Villata et al.\ (2000). Wu et al.\ (2005) found 
no apparent time lag between V and R band flux variations. Recently, Wu et al.\ (2012) monitored the source simultaneously in 
three optical bands on seven nights and found that the variability at B and V bands lagged the R band variations by 30 min on one
night.
The colour or spectral behaviour of this blazar has been studied by number of authors
(Ghisellini et al.\ 1997; Raiteri et al.\ 2003; Villata et al.\ 2004; Gu et al.\ 2006; Rani et al.\ 2010) on various timescales
but is still a subject of debate.
Some authors found a bluer-when-brighter (BWB)
chromatism (e.g., Wu et al.\ 2005; Gaur et al.\ 2012); others claimed the
opposite (e.g., Ghisellini et al.\ 1997), or no colour change with brightness even in flaring state (e.g., Stalin et al.\ 2006).
Another intriguing result pointed out by Qian et al.\ (1996) and Wagner et al.\ (1996) is that the radio spectral index of 
this source was found to correlate with the intranight optical variations, thus giving further evidence for correlated variability 
in the radio and optical ranges as noted earlier by Quirrenbach et al.\ (1991).

In order to further investigate the characteristics of the source on IDV and STV timescales, along with any spectral changes, 
 we carried out multi-band optical observations of S5 0716+714 using two telescopes in Bulgaria and one in India, during
2014 -- 2015 when the source was reported to be in an exceptionally high state (Bachev \& Strigachev 2015).
We examined colour vs magnitude correlations
that can be helpful to shed some light on the acceleration and cooling mechanisms contributing to the blazar
variability. In Section 
2 we describe the observations and data reduction procedures. Section 3 includes our approaches to microvariability analysis 
and timescale detection, while Section 4 presents our  results.
A discussion and our conclusions are presented in Section 5.



\begin{table}
\caption{ Observation log of photometric observations of S5 0716+714.}


\textwidth=8.0in
\textheight=11.0in
 \centering
\vspace*{0.2in}
\noindent

\begin{tabular}{p{1.6cm}p{1.2cm}p{1.3cm}p{1.1cm}p{1.5cm}} \hline 

 Date              & Telescope    &Number of   &Time   & Time \\
                    &              &data points &  Span        &Resolution   \\
(yyyy mm dd)         &              & (B,V,R,I) & ($\sim$ hours)  & ($\sim$ minutes)\\\hline
 2014 11 15          &A        &1,1,660,1       & 5.45 & --,--,0.4,--           \\  
 2014 11 16          &A    &  1,1,1,1       & 0.13 & --,--,--,--            \\                  
 2014 11 21          &A    &1,1,350,1       & 5.33 & --,--,0.8,--             \\                  
 2014 11 22          &A    &1,1,1,1         & 0.15 & --,--,--,--           \\                              
 2014 12 21          &A    &1,1,229,1       & 5.50 & --,--,0.8,--             \\                  
  2014 12 22          &A    &1,1,74,1       & 3.20 & --,--,0.8,--             \\                  
  2014 12 23          &A    &1,1,372,1      & 5.80 & --,--,0.8,--              \\                  
  2014 12 24          &B    &1,1,1,1        & 0.11 & --,--,--,--            \\                  
  2015 01 15          &B    &1,1,1,1        & 0.11 & --,--,--,--            \\                  
  2015 01 16          &B    &1,1,1,1        & 0.11 & --,--,--,--            \\                  
  2015 01 17         &B    &33,30,32,31     & 5.17 & 8.2,8.2,8.2,8.2                \\                  
  2015 01 18         &B    &1,1,1,1         & 0.10 & --,--,--,--            \\                  
  2015 01 19         &B    &23,1,23,24      & 4.08 & 8.2,--,8.2,8.2               \\                  
  2015 02 11         &B    &1,1,1,1         & 0.20 & --,--,--,--            \\                  
  2015 02 12         &B    &1,1,1,1         & 0.20 & --,--,--,--            \\                  
  2015 02 13         &B    &1,1,1,1         & 0.20 & --,--,--,--            \\                  
  2015 02 14         &B    &19,1,19,19      & 3.70 & 8.4,--,8.4,8.4              \\                  
  2015 02 17         &B    &1,1,1,1         & 0.10 & --,--,--,--          \\                  
  2015 02 18         &B    &1,1,1,1         & 0.10 & --,--,--,--            \\   
  2015 02 18         &C    &1,1,5,1         & 0.30 & --,--,2.2,--            \\                  
  2015 02 20         &C    &1,1,3,1         & 0.53 & --,--,12,--            \\                  
  2015 03 17         &B    &1,1,1,1         & 0.10 & --,--,--,--            \\                  
  2015 03 20         &B    &1,1,1,1         & 0.10 & --,--,--,--            \\     \hline 

     \end{tabular} 

\end{table}


\noindent

\section{\bf Observations and Data Reductions}

Our optical photometric observations of S5 0716+714 were performed in the B, V, R, and I pass-bands,
using three telescopes, two in India and one in Bulgaria, all equipped with CCD detectors. 
The details of the telescopes, instruments and other parameters used are given in Table 1.  Details 
on the dates, number of observations made in each band, total time span of observation for each night and the time resolution
in a particular filter for those nights with multiple data points are listed in Table 2. 
Light curves (LCs) displaying IDV
that covered nights when the 
observation runs were at least $\sim$~4 hrs in at least one band are displayed in Figure 1.

\subsection{\bf Optical data from Indian telescopes}

The higher cadence observations of the blazar were carried out using the 1.3-m Devasthal
 fast optical telescope (DFOT) operated by ARIES, Nainital, India.
This is a fast beam 
(f/4) telescope with a modified RC system equipped with broad band Johnson-Cousins B, V, R, I filters. DFOT 
provides a pointing accuracy better than 10 arcsec RMS (Sagar et al.\ 2011). Further details of the telescope 
are given in Table 1 (telescope A). 

The preliminary processing of the raw photometric data was carried out through standard procedures in the
IRAF\footnote{IRAF is distributed by the National Optical Astronomy Observatories, which are operated
by the Association of Universities for Research in Astronomy, Inc., under cooperative agreement with the
National Science Foundation.} software. For image pre-processing we generated a master bias frame for each
observing night which was subtracted from all twilight flat frames and all source image frames taken on that
night. Next we generated the master flat for each pass-band by median combining all the sky flat frames in that
pass-band. After that, we flat-fielded  each source
image frame to remove pixel-to-pixel inhomogeneities. Finally, we removed cosmic rays from all source image frames.

We performed  photometry of these data to find the instrumental magnitudes of the BL Lac and the comparison
stars by using the concentric circular aperture photometric technique with the Dominion Astrophysical
Observatory Photometry (DAOPHOT II) software (Stetson 1987; Stetson 1992).  We carried out aperture 
photometry with four different concentric  aperture radii, i.e., $\sim 1 \times$~FWHM, $2 \times$~FWHM, 
$3 \times$~FWHM and $4 \times$~FWHM. 
After examining the results from these different aperture radii, we observed that the best S/N was obtained 
with aperture radii = $2 \times$~FWHM, so we adopted that aperture for our final results. 
We also reduced photometry for more 
than three  stars on the same field as the source\footnote{http://www.lsw.uni-heidelberg.de/projects/extragalactic/charts/0716+714.html}.
 We then selected those two non-varying stars from the marked 1-8 stars of the above finding chart, whose
magnitudes were similar to that of the blazar to serve as comparison stars. From these two stars, 
the one with colour closer to that of the blazar was adopted as the primary standard.

\hspace*{-0.5in}
\begin{table*}
\caption{ Results of IDV observations of S5 0716+714.} 
\textwidth=7.0in
\textheight=10.0in
\vspace*{0.2in}
\noindent

\begin{tabular}{cccccccc} \hline \nonumber

 Date       & Band   &N          & F-test  &$\chi^{2}$test  &   Variable    &A\% \\
               &                     &                               &$F_{1},F_{2},F,F_{c}(0.99),F_{c}(0.999)$ &$\chi^{2}_{1},
\chi^{2}_{2},\chi^{2}_{av}, \chi^{2}_{0.99}, \chi^{2}_{0.999}$  & & \\\hline 

 15.11.2014   & R     & 660  & 28.01, 26.72, 27.36, 1.20, 1.27 & 19319, 22558, 20938.5, 746.39, 776.91  & Var  & 12.89 \\
 21.11.2014   & R     & 350  & 19.04, 19.37, 19.21, 1.28, 1.39 & 5192.1, 8315.4, 6753.7, 413.39, 436.37 & Var & 11.59 \\
 21.12.2014   & R     & 229  & 5.30, 5.18, 5.24, 1.36, 1.51 & 427.3, 482.5, 454.9, 280.6, 299.7 & Var  &  8.99 \\
 22.12.2014   & R     & 74  & 2.69, 2.49, 2.59, 1.73, 2.08 & 109.92, 150.39, 130.15, 104.01, 116.09 & Var  & 6.87 \\
 23.12.2014   & R     & 372  & 10..89, 8.39, 9.64, 1.27, 1.38  & 6225.6, 2077.3, 4151.45, 437.29, 460.90  & Var  & 14.40 \\
 17.01.2015   & B     & 33  & 10.60, 9.88, 10.24, 2.32, 3.09 & 128.23, 148.48, 138.36, 53.49, 62.49 & Var  & 22.28 \\
              & V     & 30  & 8.16, 6.43, 7.29, 2.42, 3.29  & 1540.4, 220.8, 844.44, 49.59, 58.30  & Var & 21.69 \\
              & R     & 32  & 10.69, 10.12, 10.40, 2.35, 3.15  & 2298.8, 490.04, 1394.42, 52.19, 61.10 & Var & 20.50 \\
              & I     & 31  & 10.35, 9.12, 9.74, 2.39, 3.22  & 2336.3, 374.7, 1355.5, 50.89, 59.70 & Var & 17.59 \\              
              &(B-V)     & 29  & 0.61, 1.76, 1.19, 2.46, 3.36  & 26.2, 40.5, 33.3, 48.28, 56.89  & NV & -- \\              
              &(B-I)     & 29  & 0.55, 1.93, 1.24, 2.46, 3.36  & 19.4, 34.9, 27.1, 48.28, 56.89 & NV & -- \\              
              &(V-R)     & 30  & 0.99, 1.39, 1.19, 2.42, 3.29  & 169.8, 46.8, 108.3, 49.59, 58.30 & NV & -- \\              
              &(R-I)     & 30  & 0.18, 1.43, 0.81, 2.42, 3.29  & 32.91, 20.26, 26.58, 49.59, 58.30 & NV & -- \\              
 19.01.2015   & B     & 23  & 30.60, 30.19, 30.39, 2.78, 3.98  & 216.3, 219.9, 218.1, 40.29, 48.27  & Var  & 16.85 \\
              & R     & 23  & 36.79, 33.72, 35.26, 2.78, 3.98  & 1114.3, 667.5, 890.9, 40.29, 48.27 & Var & 14.48 \\
              & I     & 24  & 24.33, 19.52, 21.92, 2.72, 3.85  & 755.2, 340.9, 548.0, 41.64, 49.73 & Var & 13.88 \\              
              &(B-I)     & 22  & 1.98, 2.85, 2.41, 2.86, 4.13  & 19.3, 24.3, 21.8, 38.9, 46.8 & NV & -- \\              
              &(R-I)     & 22  & 3.85, 6.43, 5.14, 2.86, 4.13  & 139.1, 144.2, 141.6, 38.9, 46.8 & Var & -- \\
 14.02.2015    & B     & 19  & 2.24, 2.36, 2.30, 3.13, 4.68  & 28.7, 81.8, 55.2, 34.81, 42.31  & NV  & -- \\
              & R     & 19  & 2.36, 4.83, 3.59, 3.13, 4.68 & 44.1, 115.4, 79.8, 34.81, 42.31 & PV & 6.66 \\
              & I     & 19  & 6.87, 8.84, 7.85, 3.13, 4.68 & 144.3, 246.4, 195.3, 34.81, 42.31 & Var & 6.56 \\              
              &(B-I)     & 19  & 1.12, 1.0, 1.04, 3.13, 4.68  & 14.9, 28.1, 21.5, 34.81, 42.31 & NV & -- \\              
              &(R-I)     & 19 & 0.33, 0.56, 0.44, 3.13, 4.68 & 6.3, 13.4, 9.8, 34.81, 42.31 & NV & -- \\              
\hline
\end{tabular} \\
\noindent
Var : Variable, PV : probable variable, NV : Non-Variable     \\
\end{table*}

\subsection{\bf Optical data from Bulgarian telescopes}

Photometric observations of the blazar were carried out using two telescopes in Bulgaria (B and C
in Table 1). The 50/70 cm Schmidt telescope at Rozen 
National Astronomical observatory, made observations with BV Johnson and RI Cousins filters. The 60 cm 
Cassegrain telescope of Belogradchik AO was equipped with standard UBVRI filter sets.
Instrumental details are summarized in Table 1.
 
Standard data reduction including bias subtraction, dark-current (where appropriate) and flat field 
corrections were done using the MIDAS\footnote{ESO-MIDAS is the acronym for the European Southern Observatory 
Munich Image Data Analysis System which is developed and maintained by European Southern Observatory} package. 
The aperture radius was taken to be typically 2 -- 3 times the seeing, adjusted to result in minimal errors. 
The calibrated LCs are displayed in Fig.\ 1.

\begin{figure*}
\epsfig{figure=  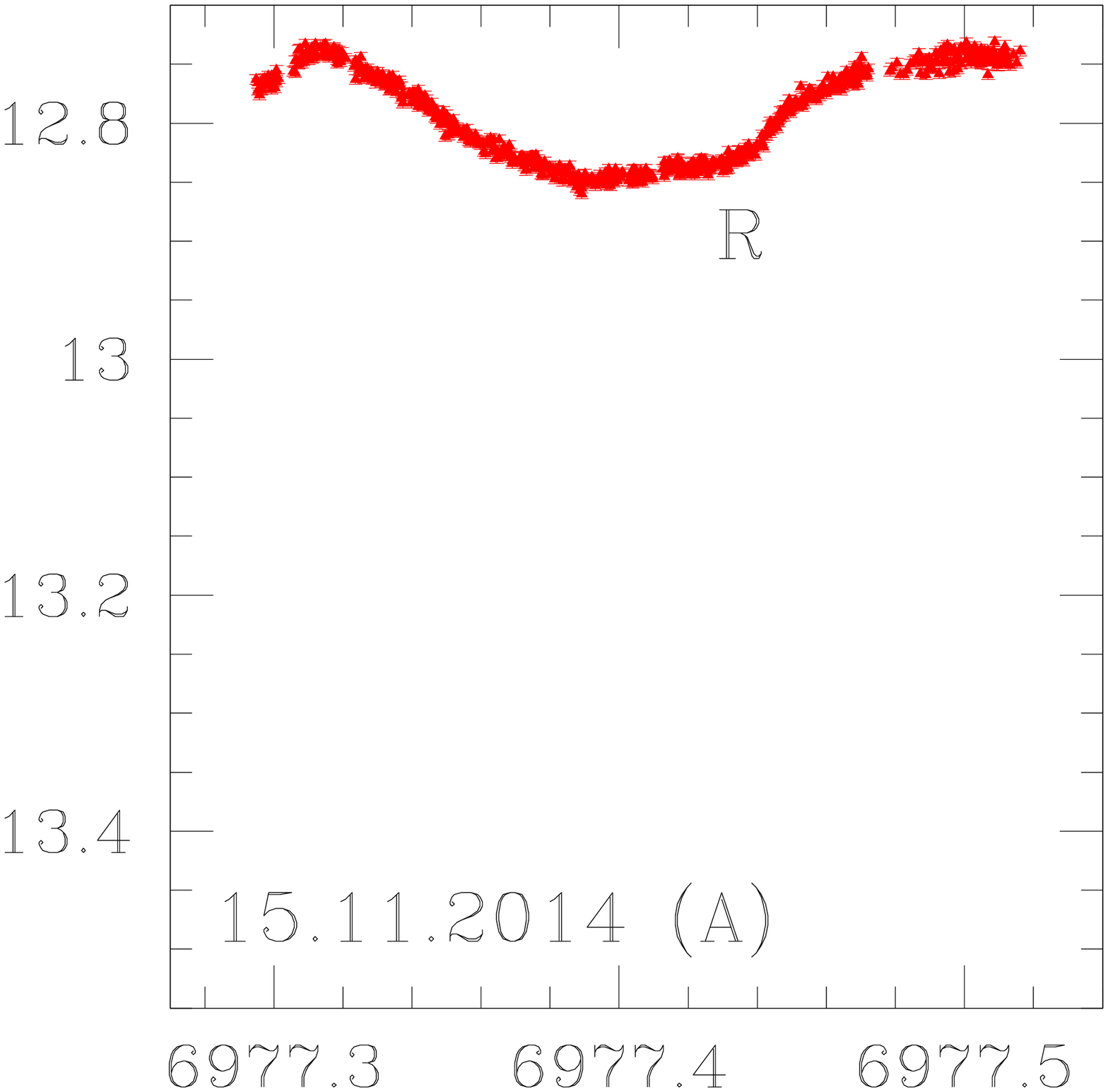,height=1.567in,width=1.59in,angle=0}
 \epsfig{figure=  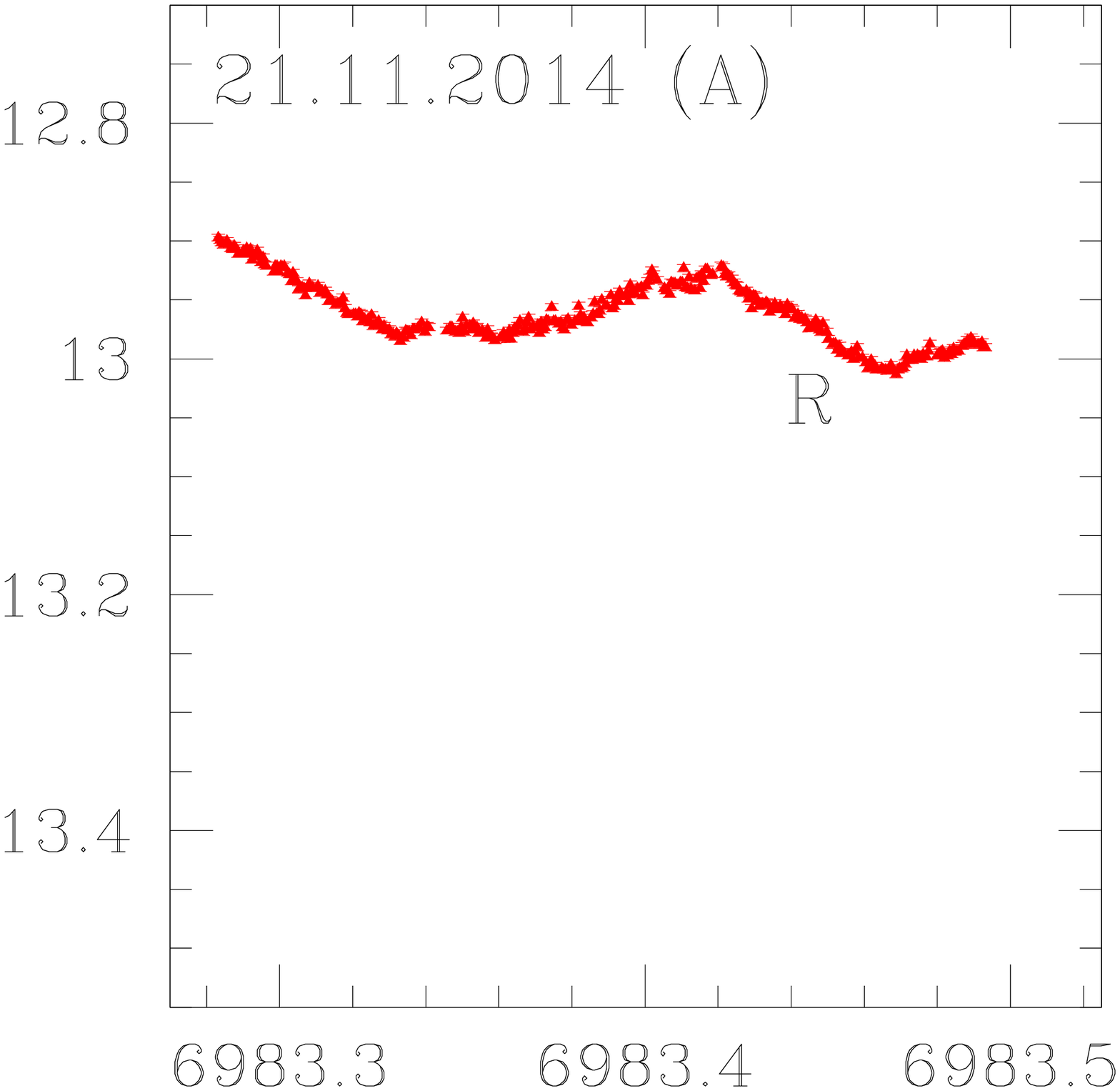,height=1.567in,width=1.59in,angle=0}
\epsfig{figure=  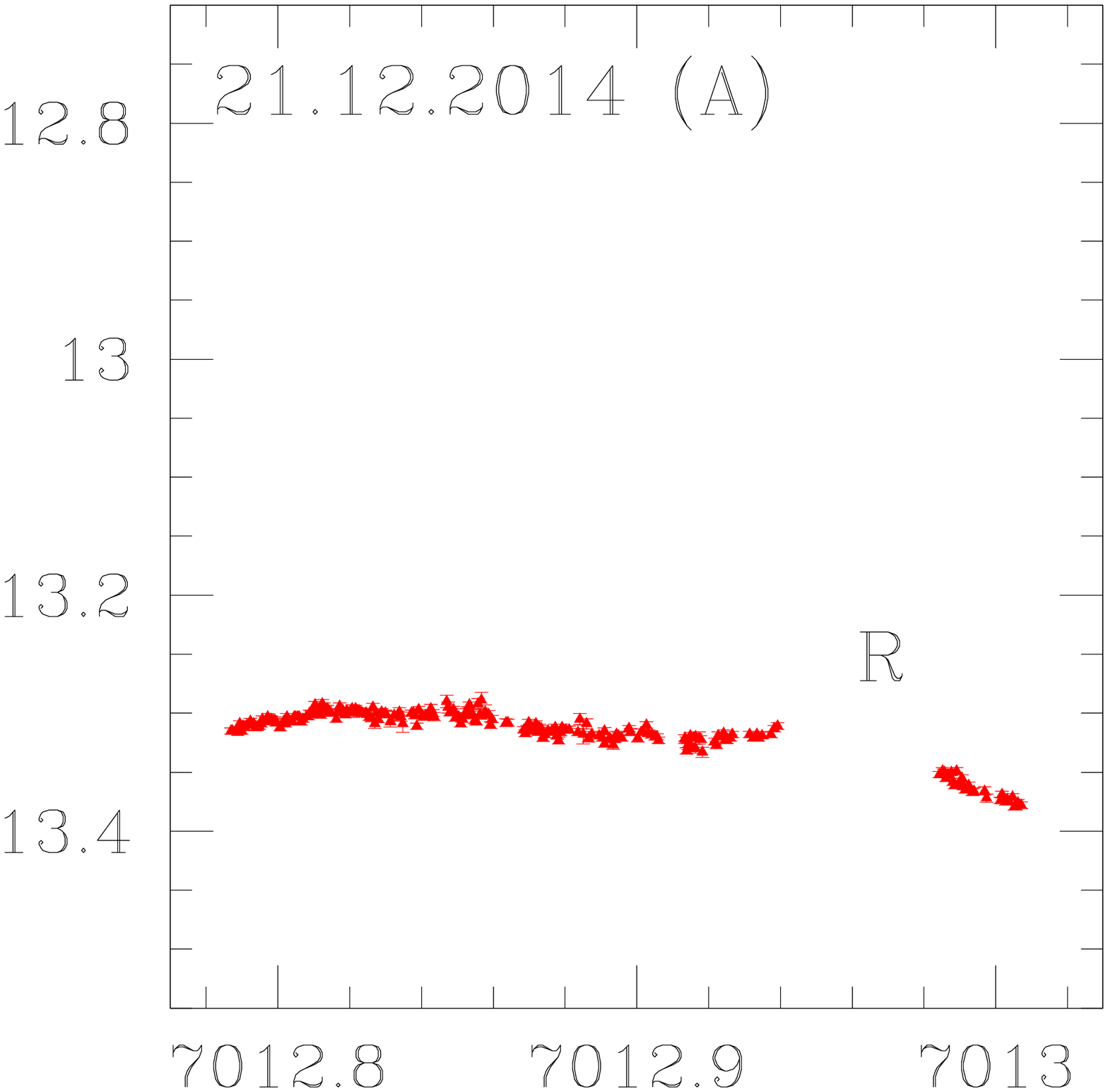,height=1.567in,width=1.59in,angle=0}
\epsfig{figure=  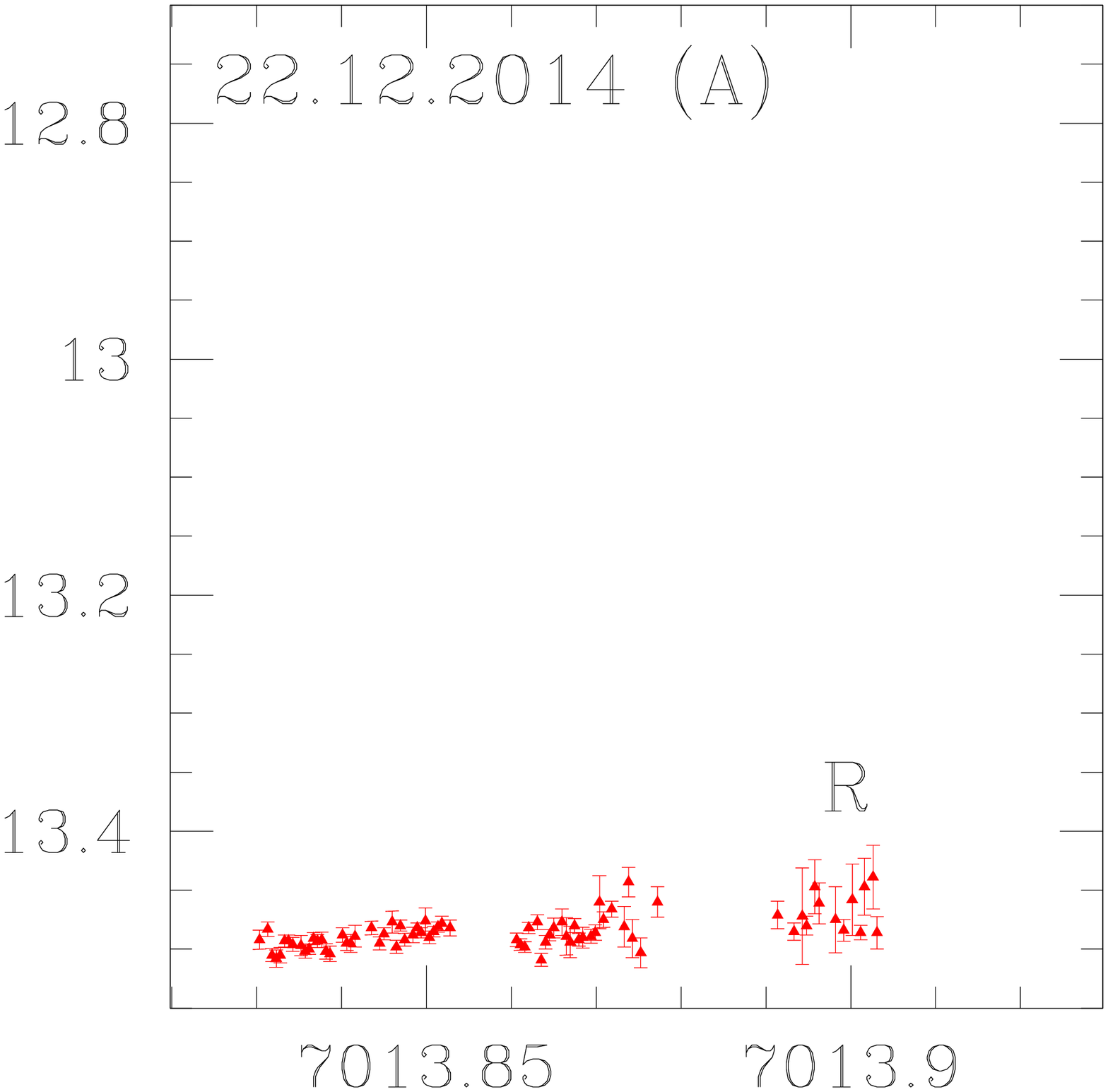,height=1.567in,width=1.59in,angle=0}
\epsfig{figure=  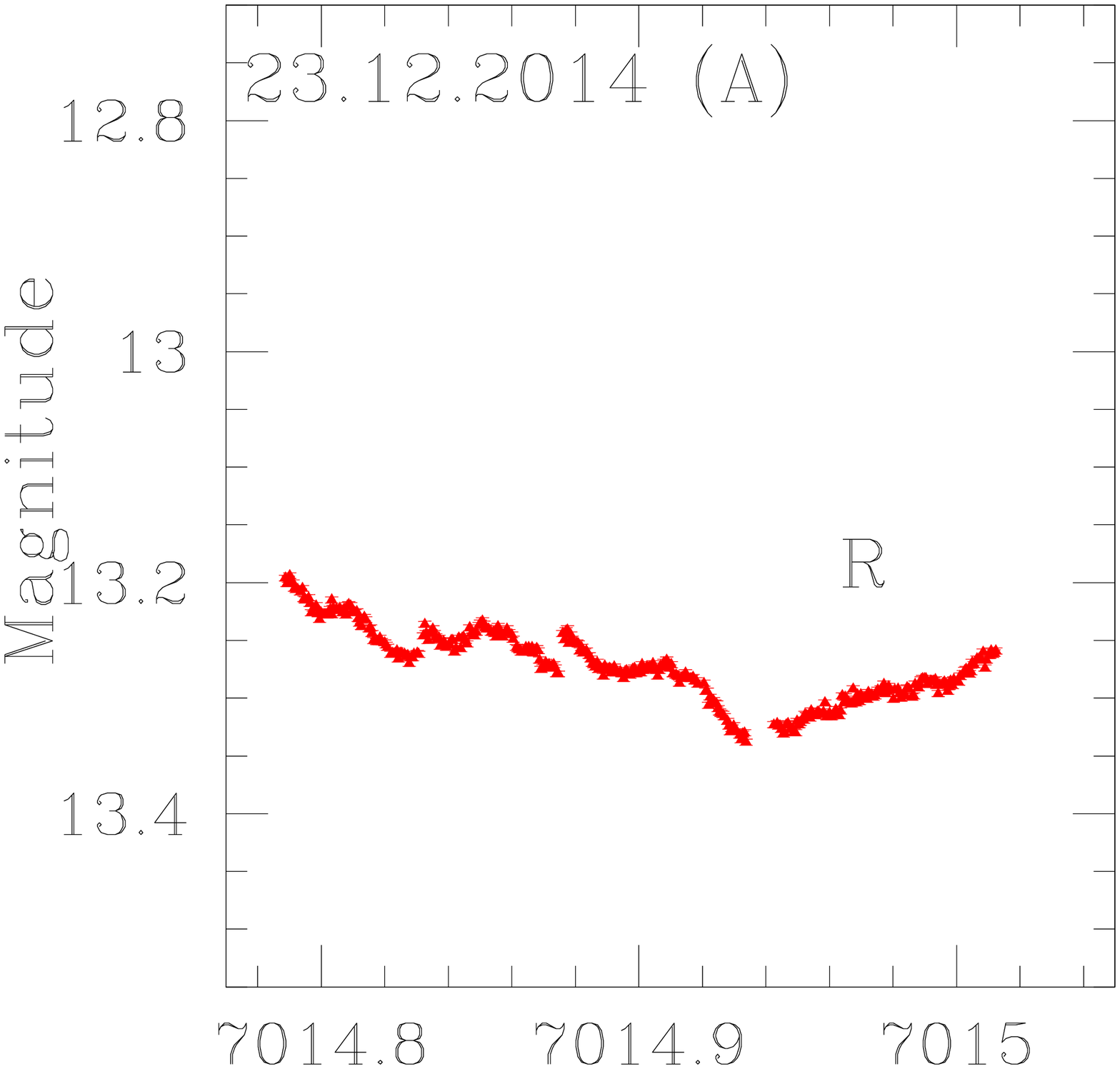,height=1.567in,width=1.59in,angle=0}
\epsfig{figure=  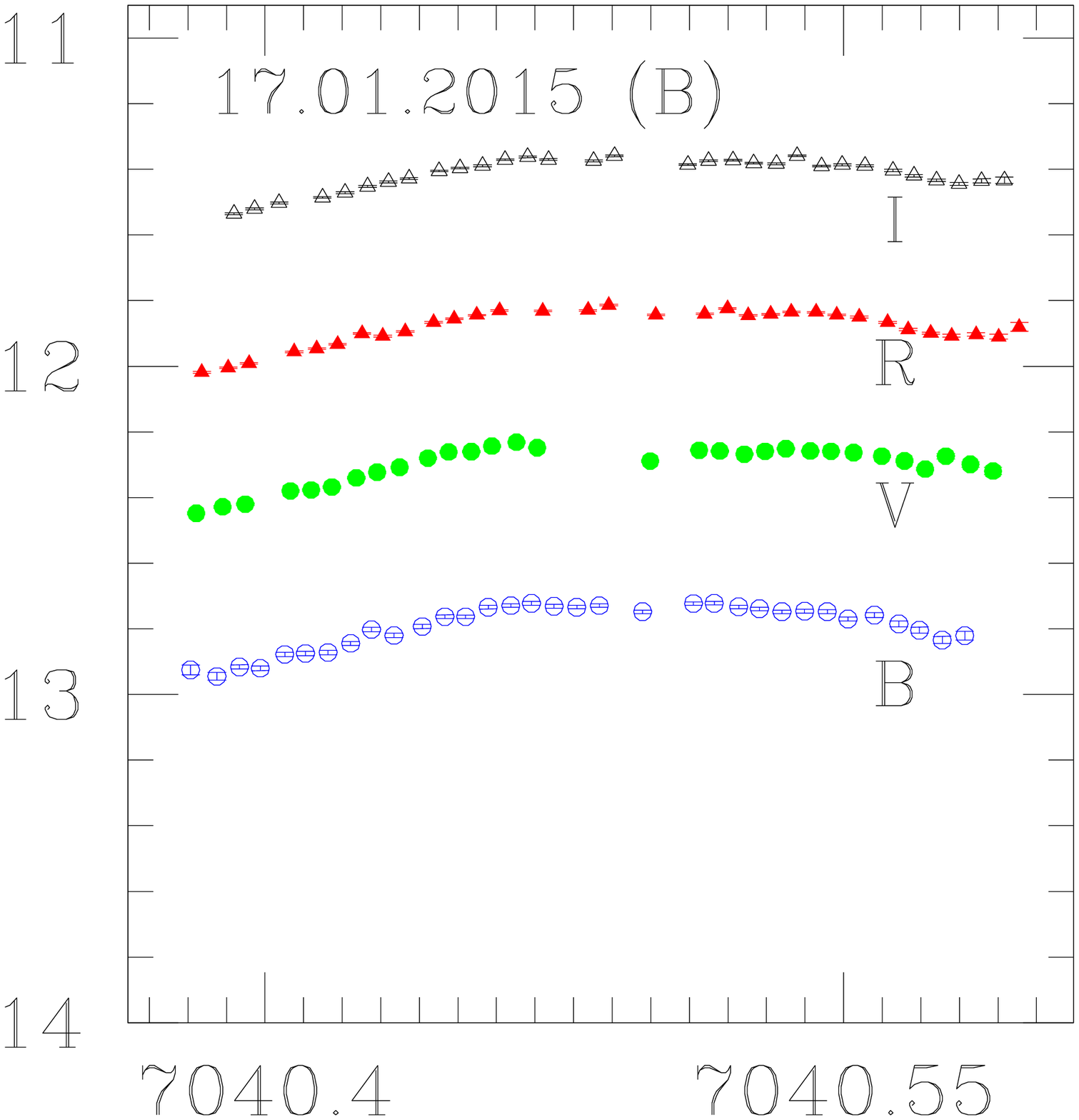,height=1.567in,width=1.59in,angle=0}
\epsfig{figure=  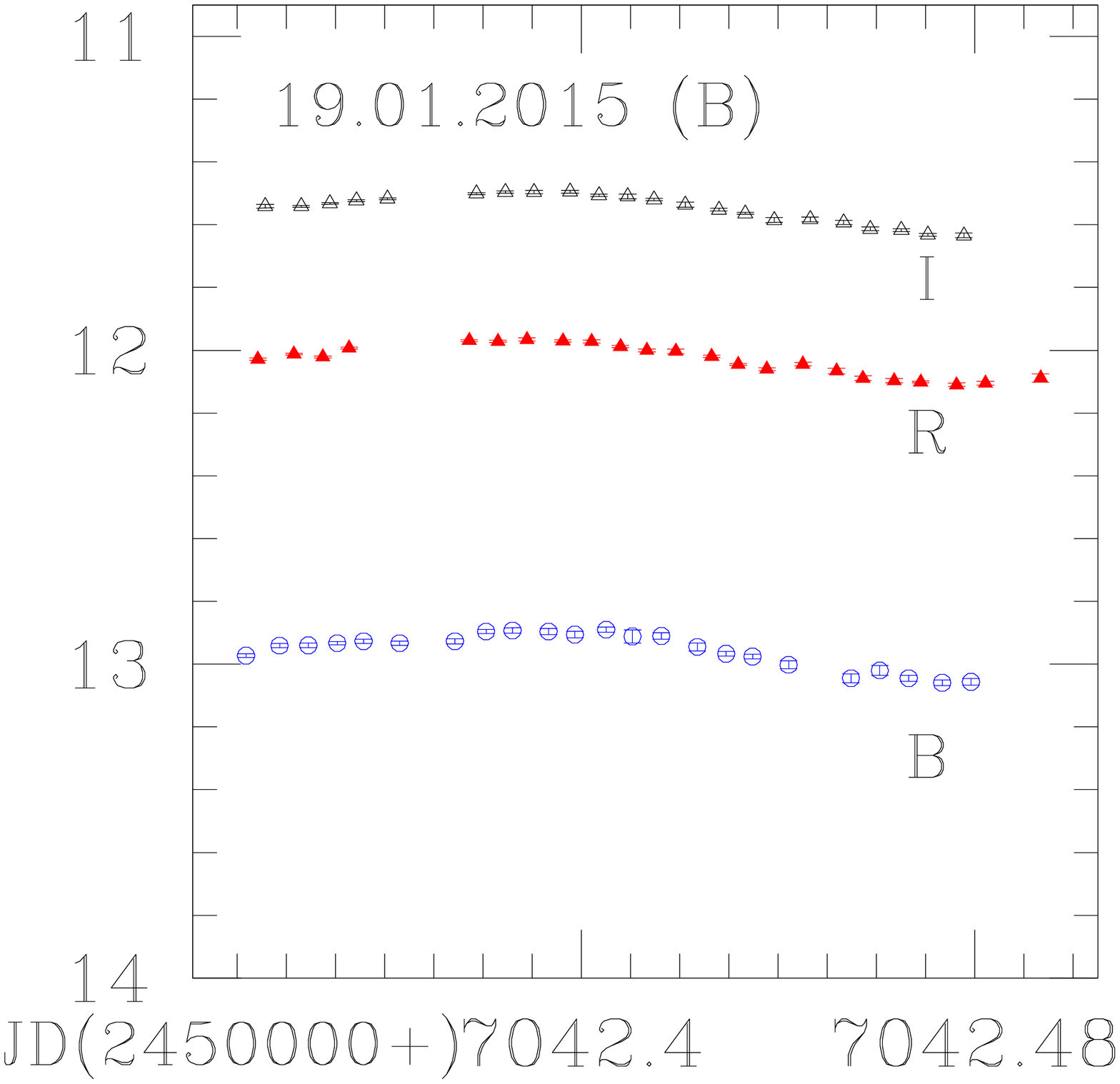,height=1.567in,width=1.59in,angle=0}
\epsfig{figure= 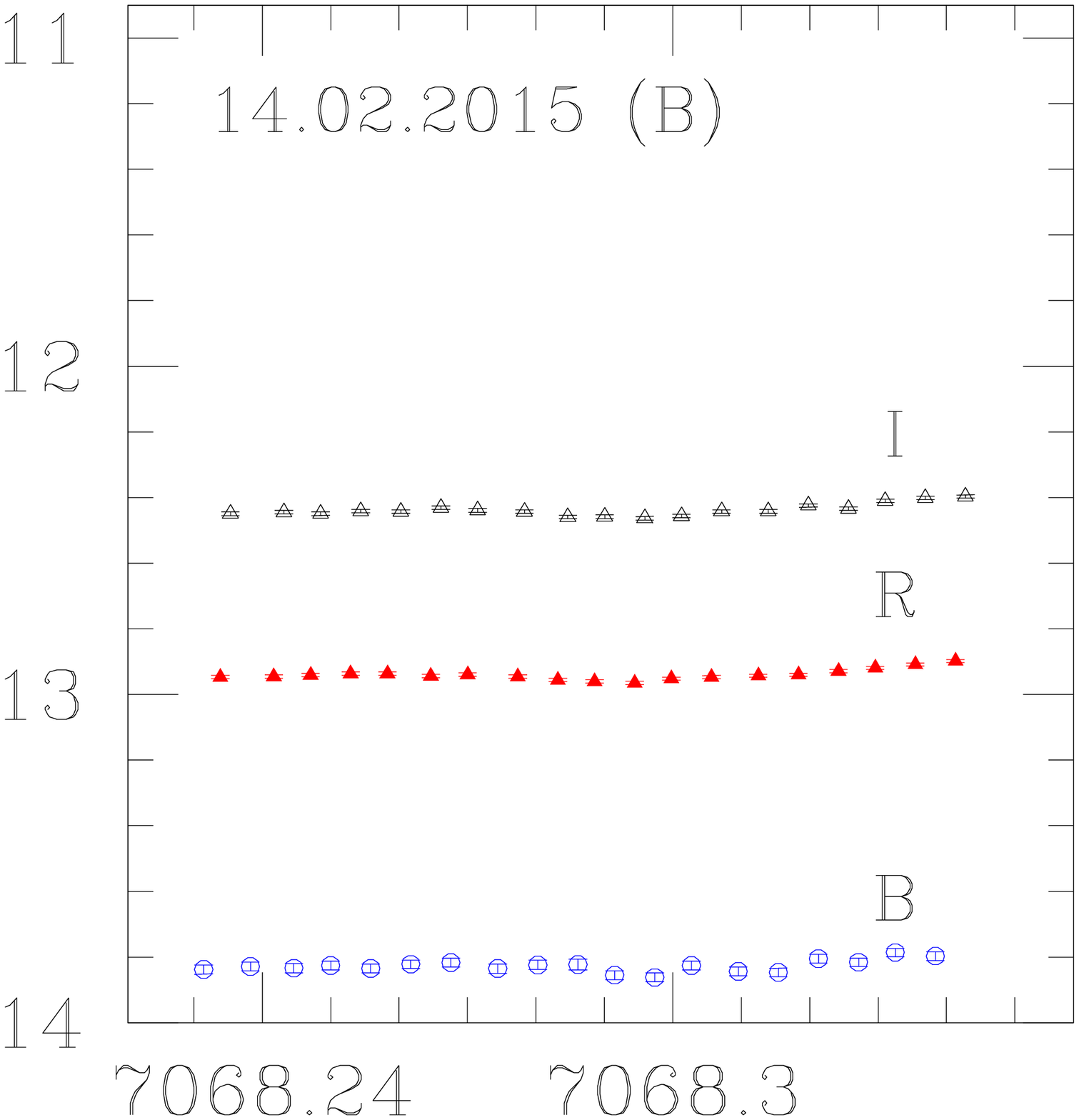,width=1.59in,height=1.567in,angle=0}
  \caption{Light curves for S5 0716+714; blue denotes B filter LC; green, V filter;  red, R filter; 
  Black, I filter. In each plot, X and Y axis are the JD and magnitude, respectively.
Observation date and the telescope used is indicated in each plot.}
\label{LC_BL}
\end{figure*}

\section{Analysis Techniques}

To search for and quantify the blazar's variability, we have employed several statistical analysis techniques. All these tools were developed
using MATLAB.

\subsection{\bf Variability detection criterion}

To quantify microvariations of the BL Lacertae object we have used two statistics, namely the F-test
and $\chi^{2}$ test. 

\subsubsection{\bf F-Test}
The F-test is considered to be a proper statistics to test
any changes of variability; $F$ values compare two sample variances and are calculated as (e.g.,  Agarwal et al.\ 2015):

\begin{equation}
 \label{eq.ftest}
 F_1=\frac{Var(BL-Star A)}{Var(Star A-Star B)}, \nonumber \\ 
 F_2=\frac{Var(BL-Star B)}{ Var(Star A-Star B)}.
\end{equation}

Here (BL-Star A), (BL-Star B), and (Star A-Star B) are the differential instrumental magnitudes of blazar and star A, blazar 
and star B, and star A and star B, respectively, while Var(BL-Star A), Var(BL-Star B), and Var(Star A-Star B) are 
the variances of those differential instrumental magnitudes.

We take the average of $F_1$ and $F_2$ to find a mean observational $F$ value.
The $F$ value is then compared with $F^{(\alpha)}_{\nu_{bl},\nu_*}$, a critical value, where $\nu_{bl}$ and $\nu_*$ 
respectively denote the number of degrees of freedom for the blazar and star, while $\alpha$ is the significance level set as
0.1 and 1 percent (i.e.\ $3 \sigma$ and $2.6 \sigma$) for our analysis. If the mean F value is 
larger than the critical value, the null hypothesis (i.e., that of no variability) is discarded.

\subsubsection{\bf $\chi^{2}$-test}

To investigate the presence of IDV we also performed a $\chi^{2}$-test. The $\chi^{2}$ statistic is defined as (Agarwal \& Gupta 2015):

\begin{equation}
\chi^2 = \sum_{i=1}^N \frac{(V_i - \overline{V})^2}{\sigma_i^2},
\end{equation}
where, $\overline{V}$ is the mean magnitude, and the $i$th observation yields a magnitude $V_i$
with a corresponding standard error $\sigma_i$ which is due to photon noise from the source and sky, CCD read-out and other non-systematic
error sources. 
Exact quantification of such errors by the IRAF reduction package is impractical and it has been found that
theoretical errors are smaller than the real
errors by a factor of 1.3-1.75 (e.g., Gopal-Krishna et al.\ 2003) which for our data is $\sim$1.5, on average. 
So the errors obtained after data reduction 
should be multiplied by this factor to get better estimates of the real photometric errors.
This statistic is then compared with a critical value $\chi_{\alpha,\nu}^2$ where $\alpha$ is again the significance level  as the in case of
the F-test while $\nu = N -1$ is the degree of freedom; $\chi^2 > \chi_{\alpha,\nu}^2$ implies the presence of variability.

\noindent

\subsubsection{\bf Percentage amplitude variation}

The percentage variation on a given night is calculated by using the variability amplitude parameter $A$,
introduced by Heidt \& Wagner (1996), and defined as
\begin{eqnarray}
A = 100\times \sqrt{{(A_{max}-A_{min}})^2 - 2\sigma^2}(\%) .
\end{eqnarray}
Here $A_{max}$ and $A_{min}$ are the maximum and minimum values in the calibrated LCs of the blazar, and $\sigma$
is the average measurement error.

\subsection{\bf Structure Function analysis}

The first order Structure Function (SF) is a very useful tool designed to search, among other things, for periodicity and timescales, thus providing some information
on the nature of physical process causing the observed variability. It was introduced in the field of astronomy by Simonetti, Cordes, 
\& Heeschen (1985) and can be applied to an unevenly sampled data series. For details about how we used the SF, see Gaur et al.\ (2010).

Emmanoulopoulos, McHardy, \& Uttley (2010) pointed out that the SF can sometimes lead to incorrect claims of periodicities or timescales.
Hence we also examined the data for timescales and possible periodicities by the
Discrete Correlation Function (DCF) method.
\noindent
\subsection{\bf Discrete correlation function (DCF)}

To quantify correlation and search for possible time lags between different optical bands we computed the DCF,
which was introduced by Edelson \& Krolik (1988) and was later generalized by Hufnagel \& Bregman (1992) for better error estimation.

 The first step is to calculate the unbinned correlation (UDCF).
For each pair of data \( (x_i , y_j ) \), with 
$0 \leq i,j \leq N$, with $N$  the number of data points, the UDCF is
\begin{eqnarray}
 UDCF_{ij}(\tau) = \frac{(x_i-\bar{x})(y_j-\bar{y})}{\sqrt{(\sigma_{x^2} - e_{x^2})(\sigma_{y^2} - e_{y^2})}}
\end{eqnarray}
where $\bar{x}, \bar{y}$ are the mean values of the two discrete data series \( x_i , y_j \),
with standard deviations \( \sigma_x\), \( \sigma_y \) and measurement errors $e_x$, $e_y$.
The DCF can be calculated by averaging the UDCF values ($M$ in number) for each time delay 
\( \Delta t_{ij} = (t_{yj}-t_{xi}) \) lying in the range  \( \tau - \frac{\Delta\tau}{2} \leq t_{ij} 
\leq \tau+ \frac{\Delta\tau}{2} \) and is expressed as
\begin{equation}
 DCF{(\tau)}= \frac{\sum_{k=1}^m UDCF_{k}}{M} ,
\end{equation}
where $\tau$ is the center of the bin of size $\Delta \tau$.
The DCF technique was proposed to work for unevenly sampled data without interpolating in the temporal domain, thus giving meaningful errors given
as:
\begin{equation}
 \sigma_{DCF(\tau)} = \frac{\sqrt{\sum_{k=1}^{M} (UDCF_k-DCF(\tau))^2}}{M-1} .
\end{equation}
When the same data train is used, i.e., $(x=y)$, it is called
the Auto Correlation Function (ACF) and has
a peak at zero lag, indicating that there is no time lag between the two,
but any other strong peak indicates the presence of periodicity.
DCF value $> 0$ 
implies that two data signals are correlated, while the two anti-correlated ones have a DCF $<$ 0, and a DCF 
value equal to 0 implies no correlation between the two data trains.
For details on how we used the DCF see Hovatta et al.\ (2007); Rani, Wiita, \& Gupta (2009);
Wu et al.\ (2012); Agarwal \& Gupta (2015) and references therein.

\begin{figure}
\epsfig{figure=  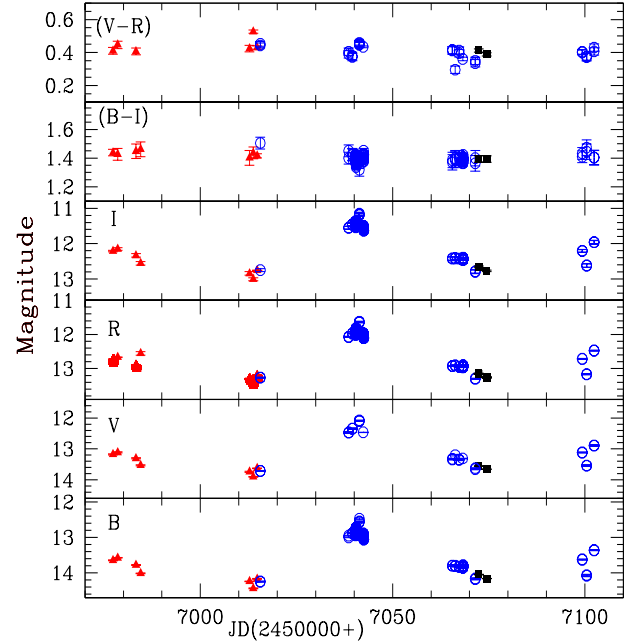,height=3.7in,width=3.3in,angle=0}
\caption{Light curves for S5 0716+71 covering full monitoring period; 
Different colours denote data from different observatories: red, telescope A; blue denotes telescope B; black denotes telescope C.
Y axis is the magnitude while x-axis is Julian Date (JD).
Top two panels represent (V-R) and (B-I) colour variation.}
\end{figure}

\begin{figure*}
\epsfig{figure=  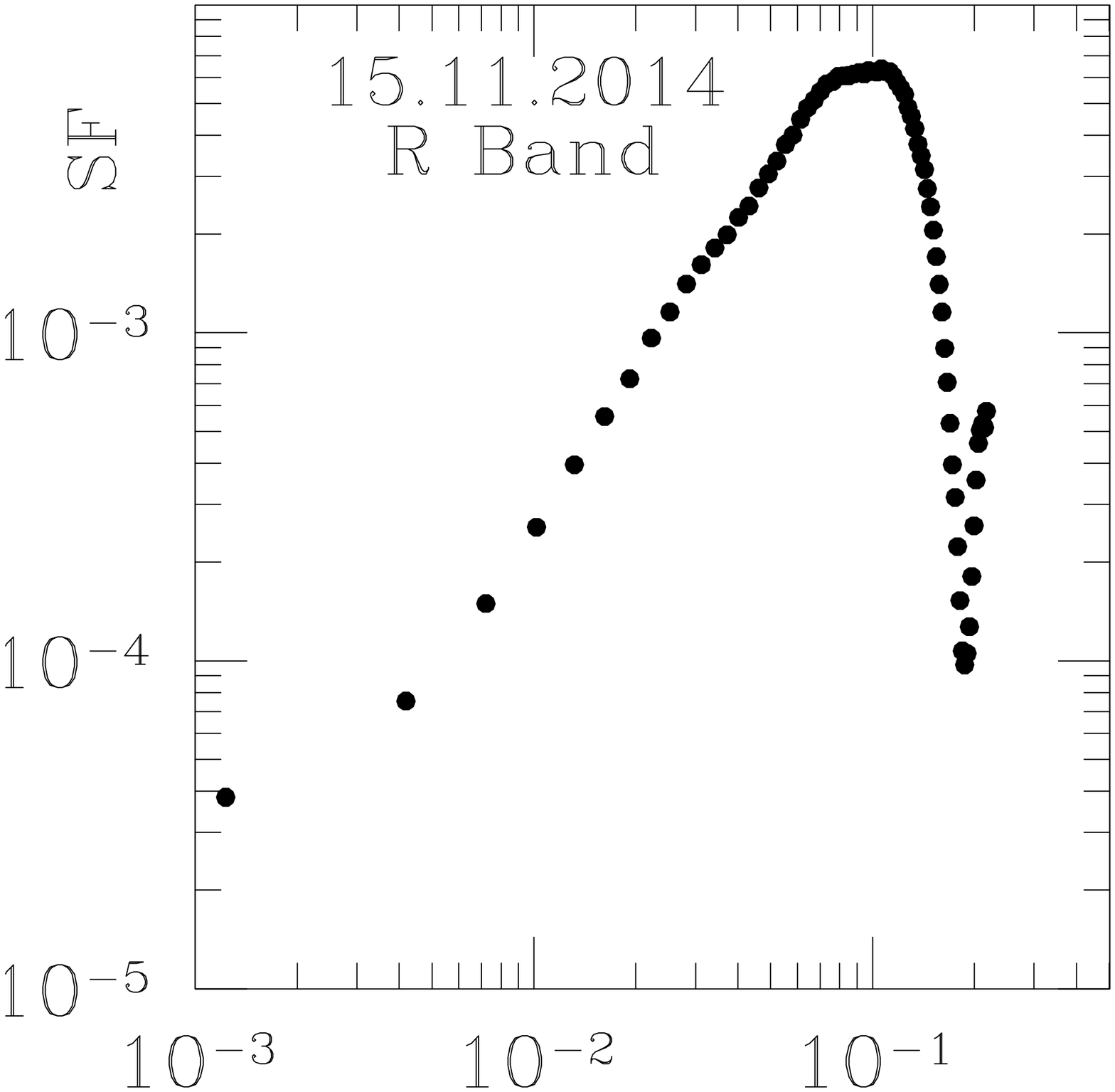,height=1.62in,width=2.1in,angle=0}
\epsfig{figure=  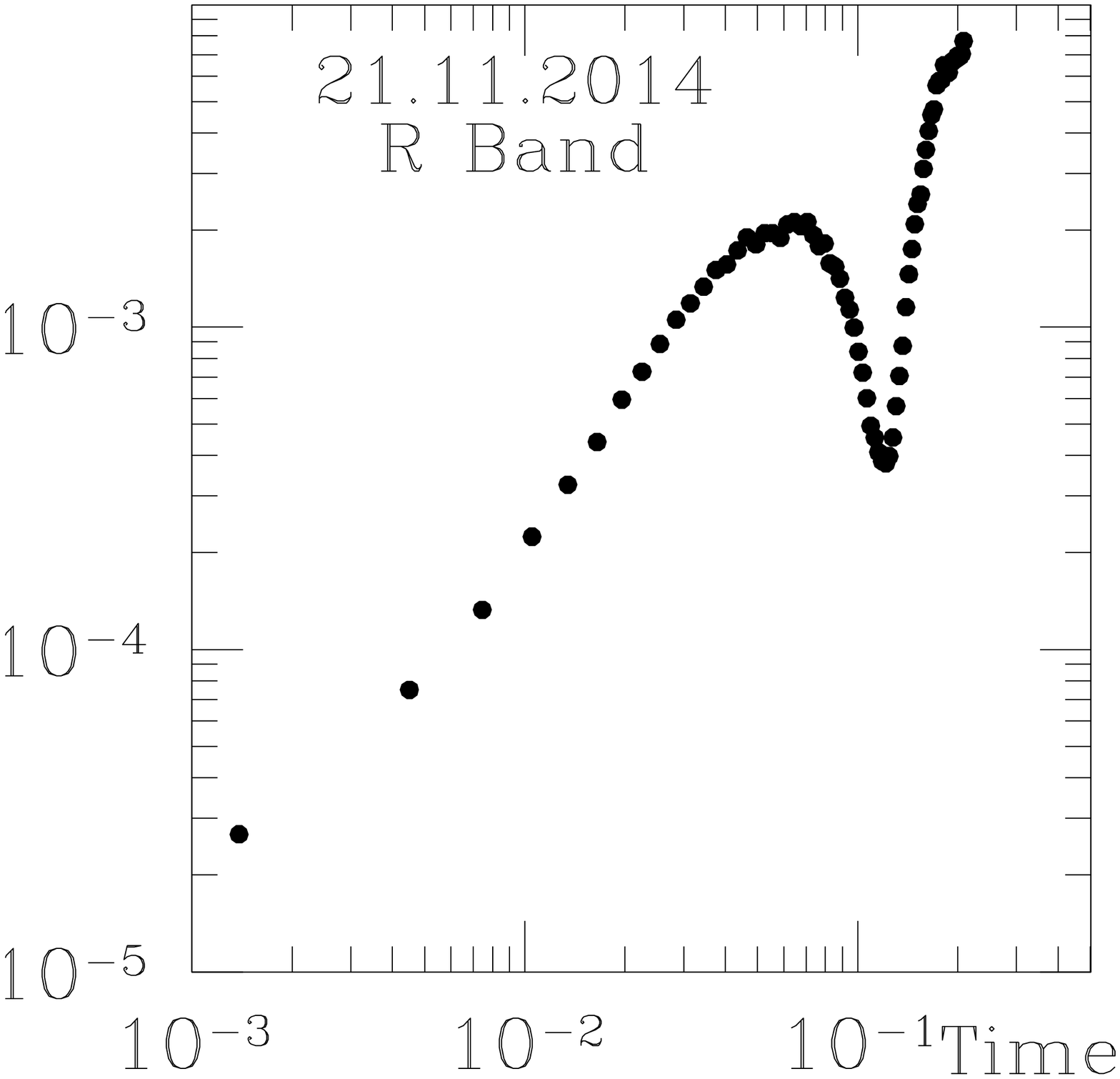,height=1.62in,width=2.1in,angle=0}
\epsfig{figure=  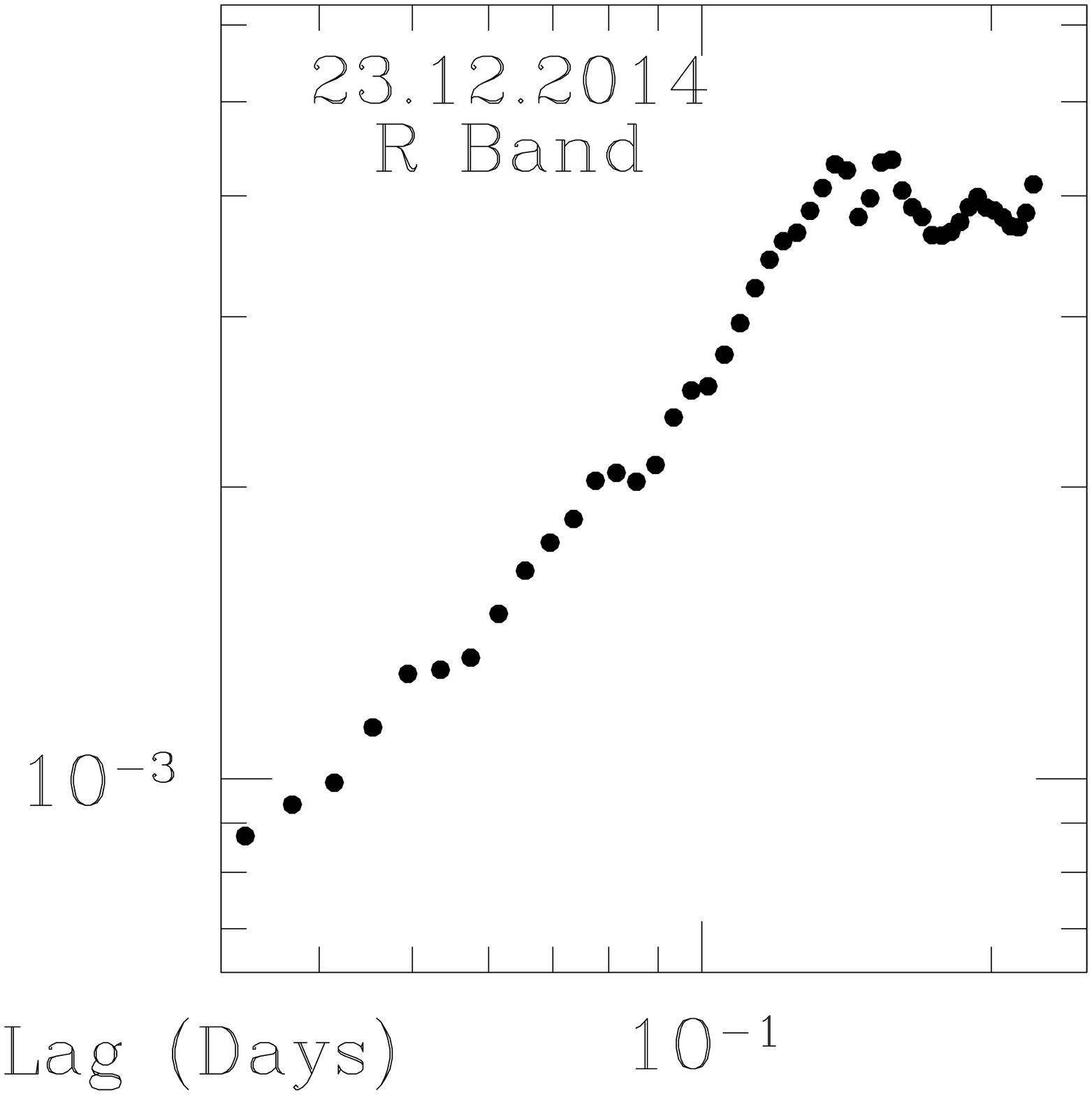,height=1.62in,width=2.1in,angle=0}
\caption{SF plots for the blazar S5 0716+714. Observation date and the filter used are indicated in each plot.}

\end{figure*}

\section{\bf  Results}

\subsection {Flux and colour variability}

 We observed the BL Lac object
S5 0716$+$714 quasi-simultaneously in the B, V, R and I passbands for a
total of 23 nights during Nov 2014 -- March 2015. The
complete observation log is given in Table 2. The IDV LCs are plotted in Fig.\ 1 and the STV plots are displayed in Fig.\ 2.
To quantify microvariability we applied both F- and $\chi^{2}$-tests, the results of which are presented
in Table 3.
The blazar is marked as variable (Var) if the variability conditions for both tests are satisfied  for the 0.999 level, while it is marked
probably variable (PV) if conditions for either of the two tests are met at the 0.99 level; the LC or colour index is marked mon-variable (NV) if none of
these conditions are met.

The source remained active during the entire monitoring period. The source was found to be variable on 7 out of a total of 8 nights in the
R filter. While it was found to display a good likelihood of variability on the other night, it did not satisfy the F-test at the 0.999 level, and so was designated PV. In the
B passband the source was observed for a total of
3 nights and was found to display clear variability on 2 of these nights. In the V and I bands the source was observed on 1 and 3 nights, respectively,
and was found to be variable during all of them. The above results are summarized in Table 3. The maximum amplitude of variability reached 22.28\%
in the B passband on 2015 Jan 01. A prominent flare was seen on 2014 Nov 15 with a magnitude change of about $\sim$ 0.13, as is evident from the IDV LC
in Fig.\ 1. As clear from Fig.\ 2, the source was observed to decline in luminosity during the end of 2014, followed by a large
increase in the source brightness during early 2015, reaching an R magnitude of 11.6 on 18 Jan,  which is the brightest  value ever recorded
for S5 0716$+$714. The apparent magnitudes in other bands on the same night were: B = 12.47, V = 12.07, and I = 11.15. Real STV was clearly seen as the
source displayed a magnitude change of 1.9 in the R passband, with the maximum brightness  of 11.6 mag on JD 2457041.3428 and
a minimum of 13.5 mag on 2457013.8639.  A mild trend of larger variability amplitude as the source gets brighter was found for the intranight variations.

We also have studied (B-R), (R-I), and (B-I) colour indices on an intraday basis using the same analysis criterion as for IDV analysis and found that
the source showed clear colour variations  on only 1 night, i.e., 2015 Jan 19 in (R-I) colour.
The behaviours of (V-R) and (B-I) on the longer
STV basis are displayed in Fig.\ 2 top two panels. The maximum (V-R) colour variation in the source on short timescales was found to be $\sim$ 
0.7 (between its colour index of $\sim$ 0.98 on JD 2456984.4618 and $\sim$ 0.30 on 2457066.2717)
while for (B-I) maximum colour index variation is found to be $\sim$ 0.2
(between $\sim$ 1.50 on JD 2457015.5914 and $\sim$ 1.32 on JD 2457041.3457).
Larger values of (B-I) as compared to (V-R) are expected, as the variances increase with frequency
separation among two bands. 

\begin{figure*}
\epsfig{figure=  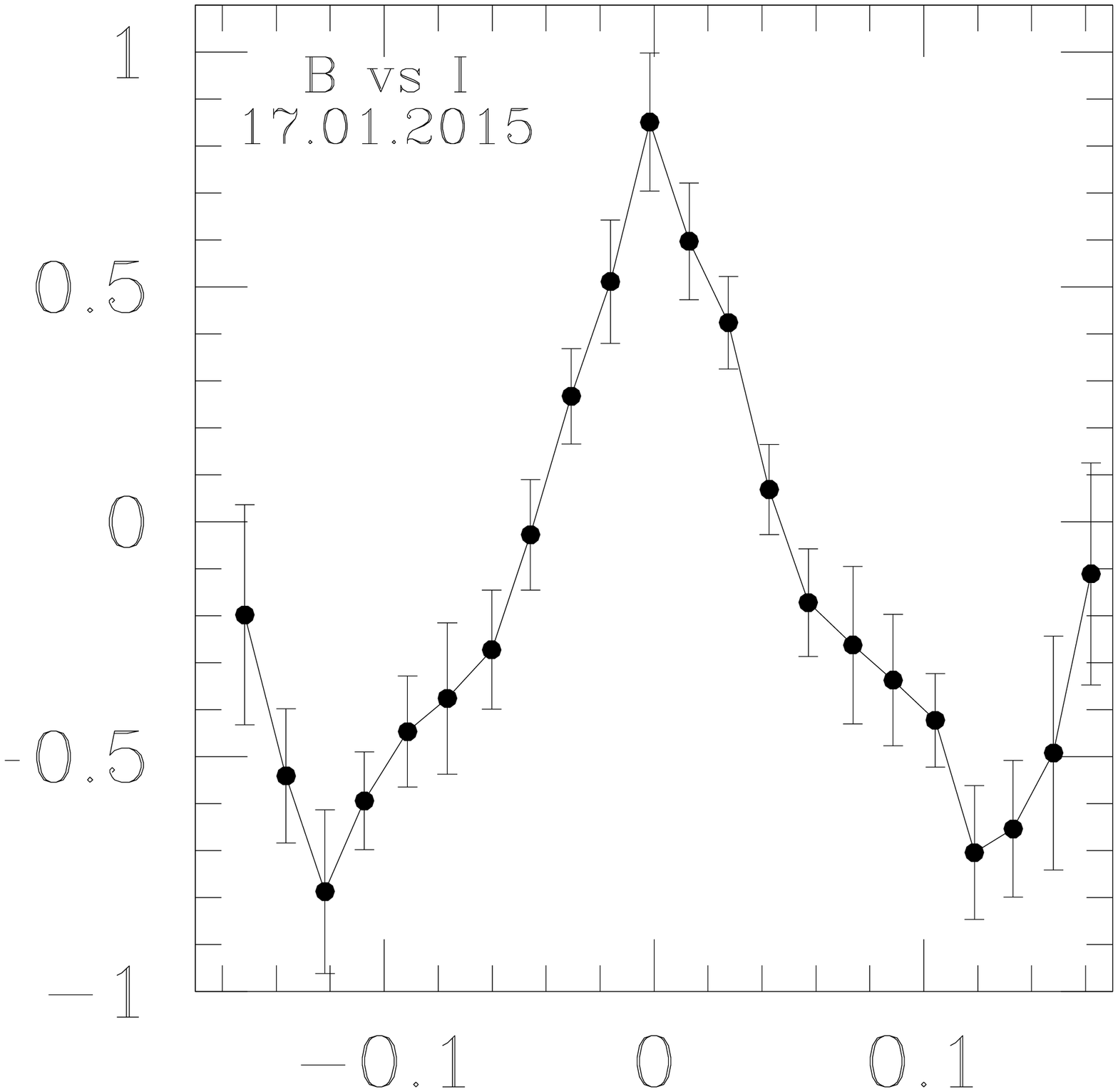,height=1.567in,width=1.59in,angle=0}
\epsfig{figure=  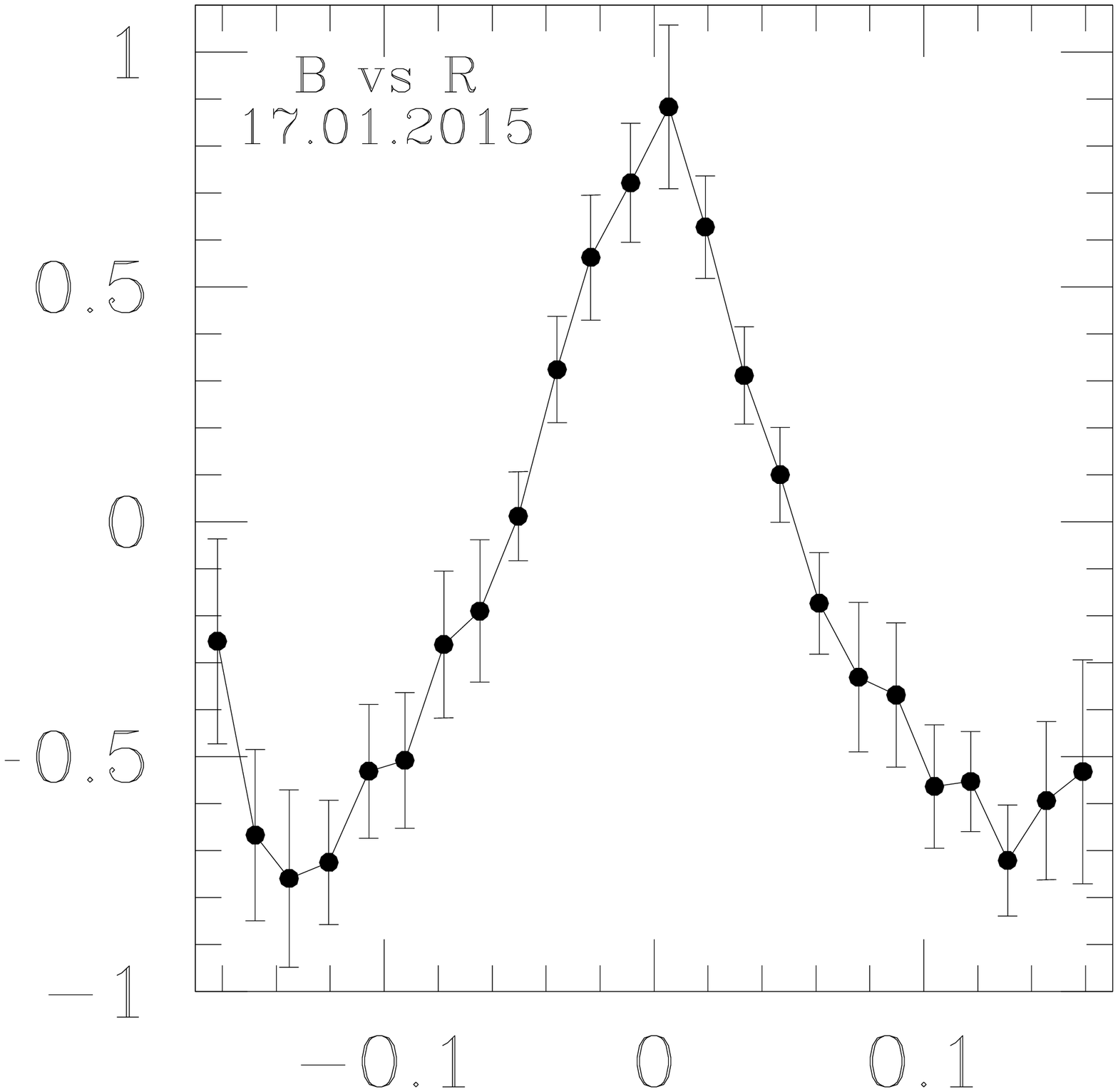,height=1.567in,width=1.59in,angle=0}
\epsfig{figure=  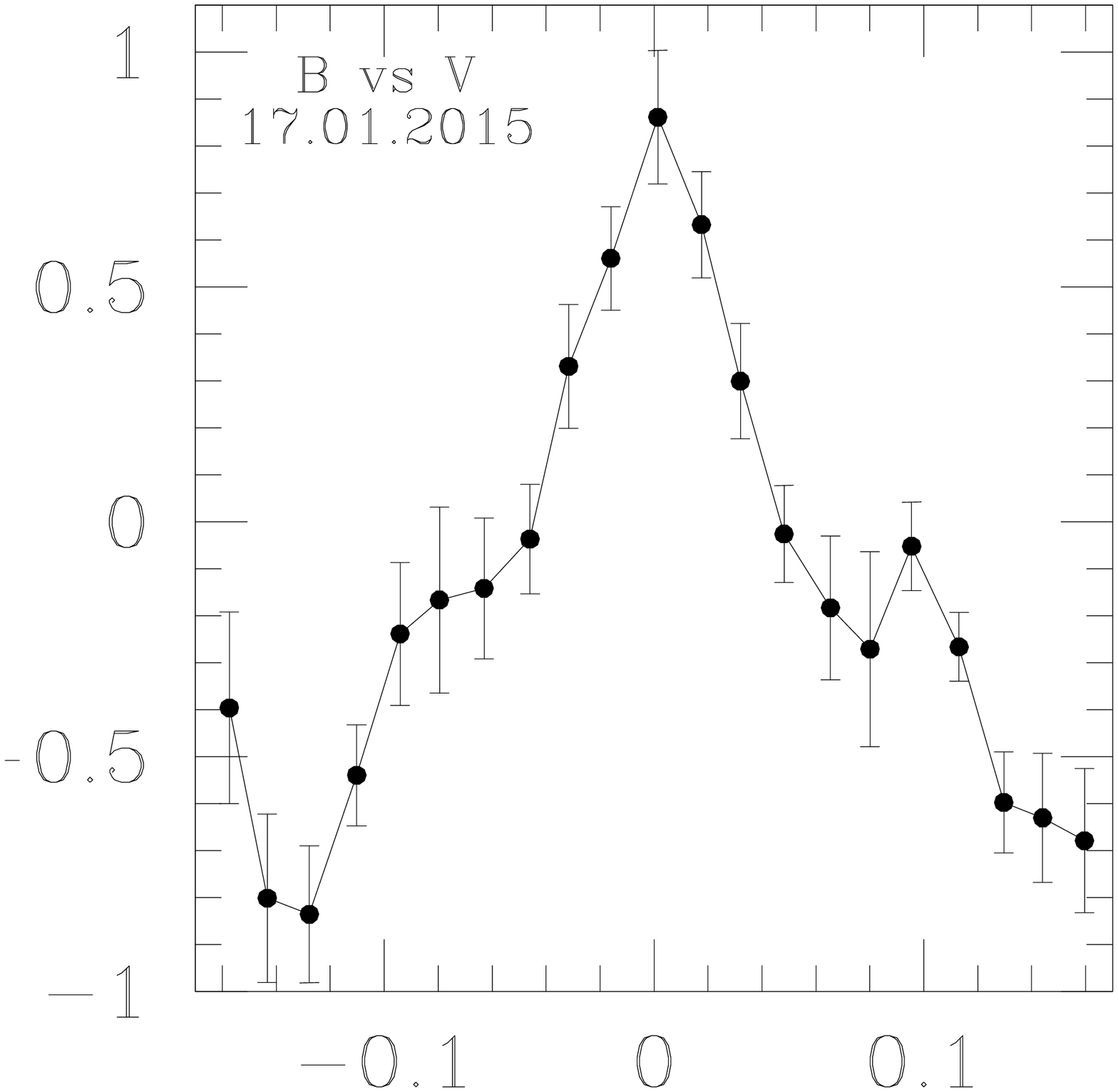,height=1.567in,width=1.59in,angle=0}
\epsfig{figure=  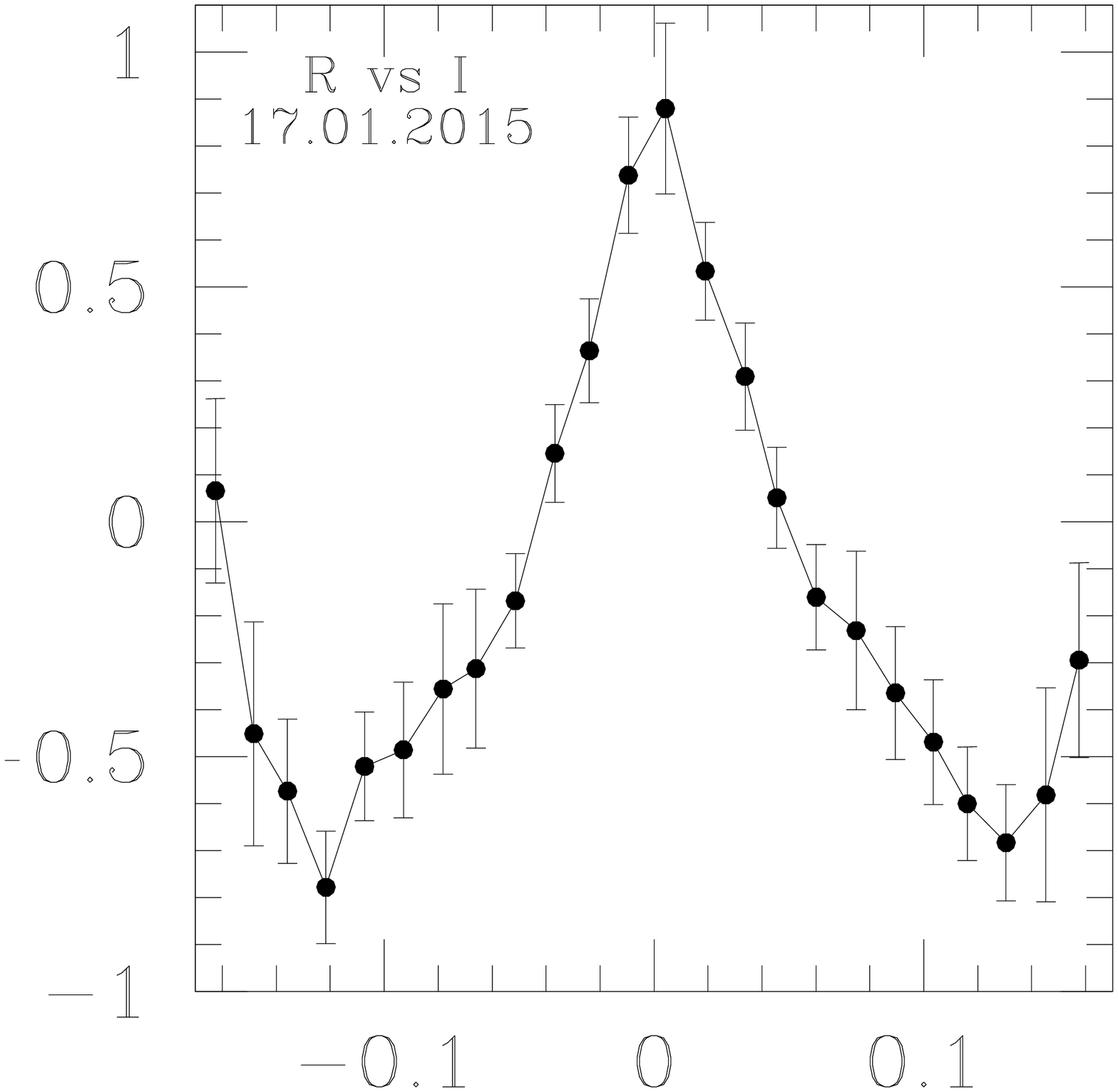,height=1.567in,width=1.59in,angle=0}
\epsfig{figure= 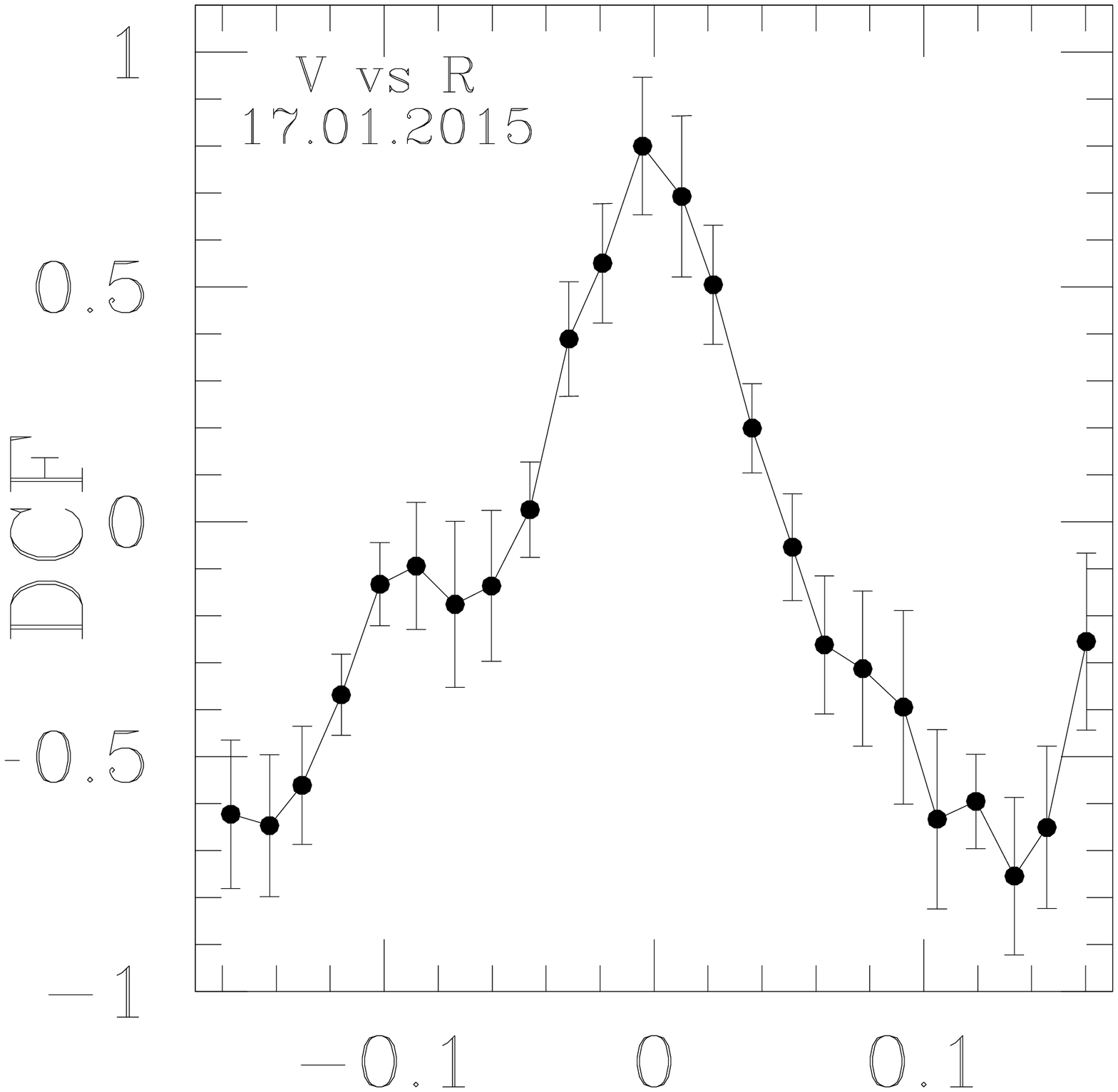,width=1.59in,height=1.567in,angle=0}
\epsfig{figure= 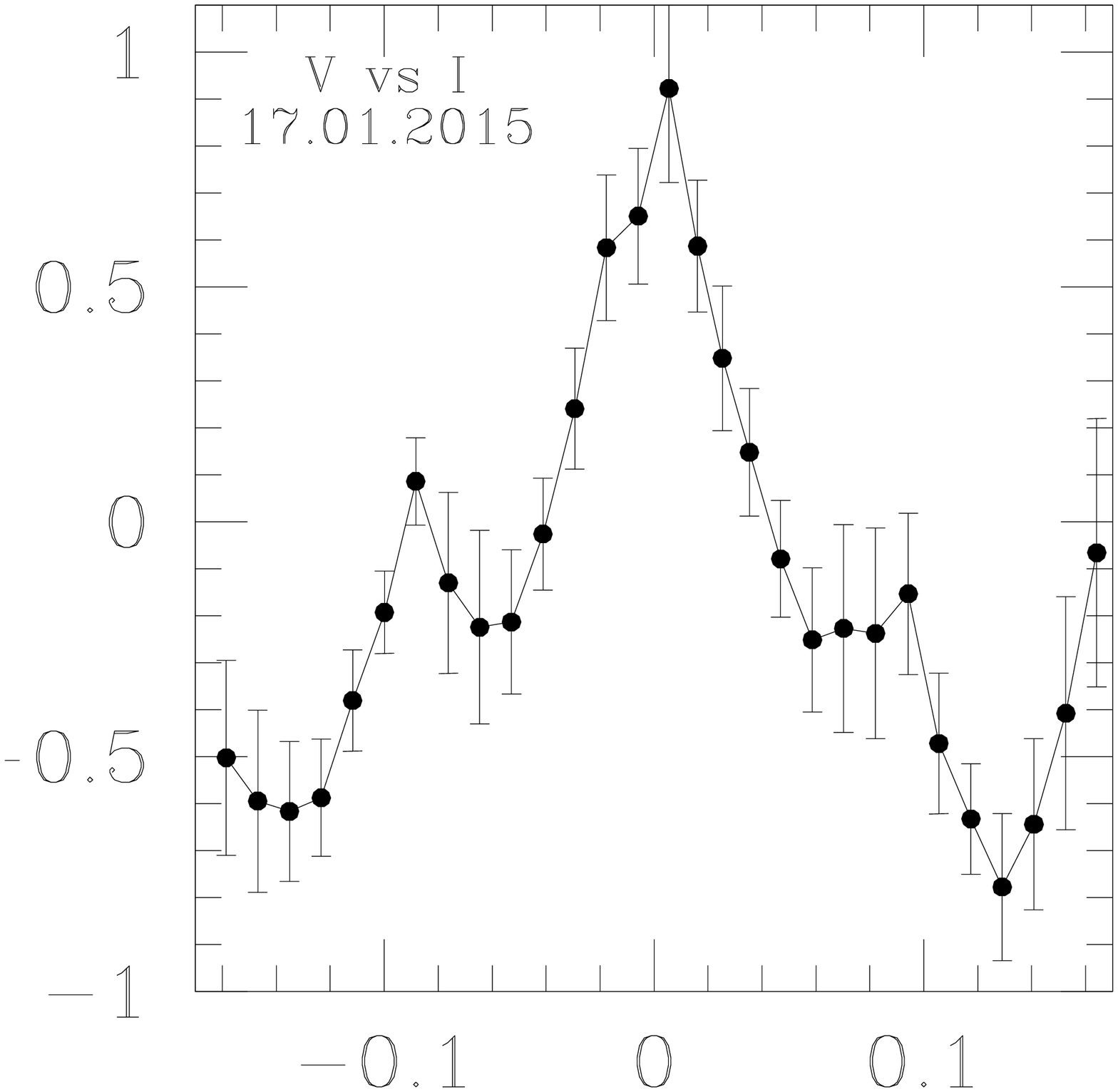,width=1.59in,height=1.567in,angle=0}
\epsfig{figure=  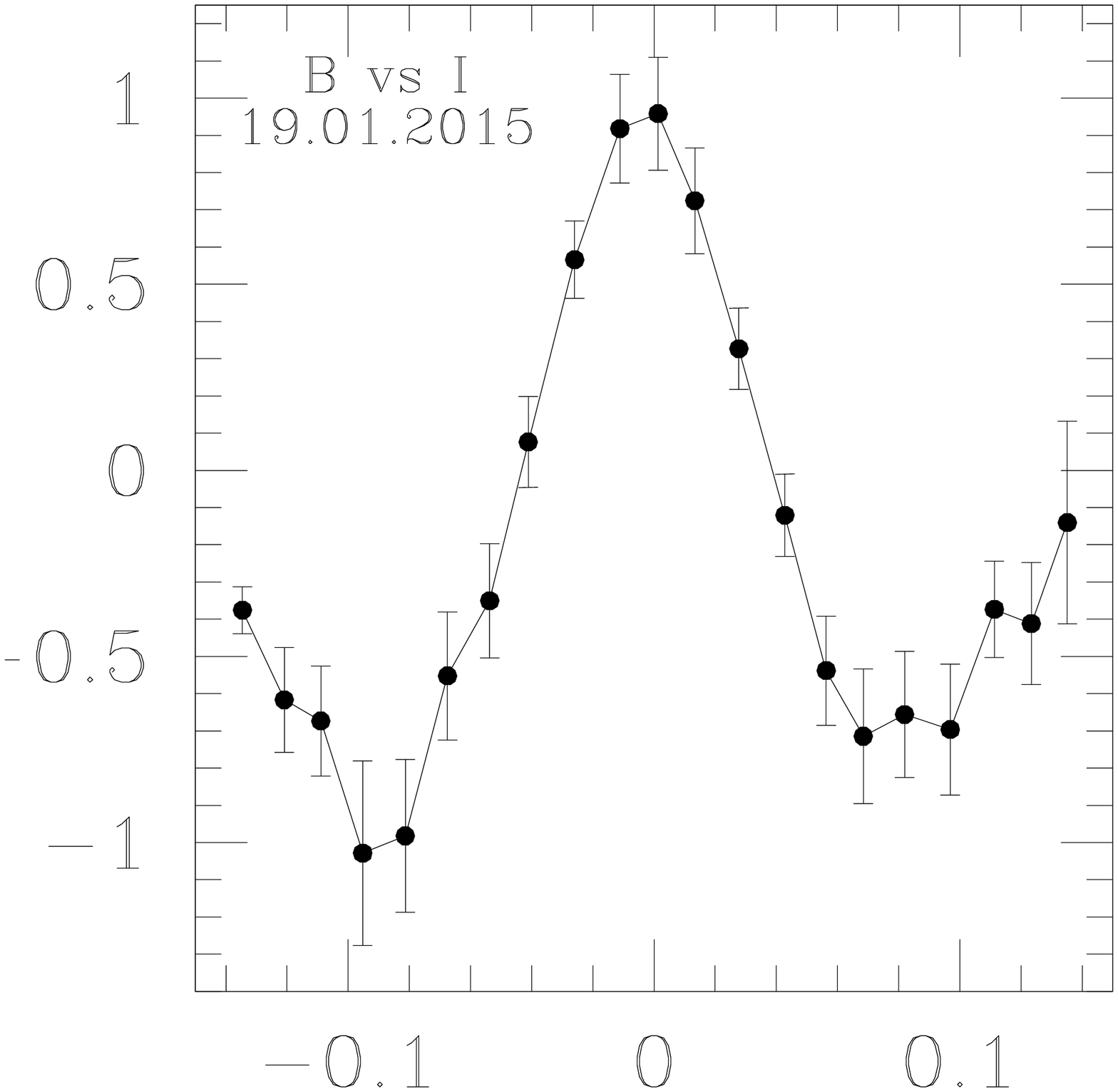,height=1.567in,width=1.59in,angle=0}
 \epsfig{figure=  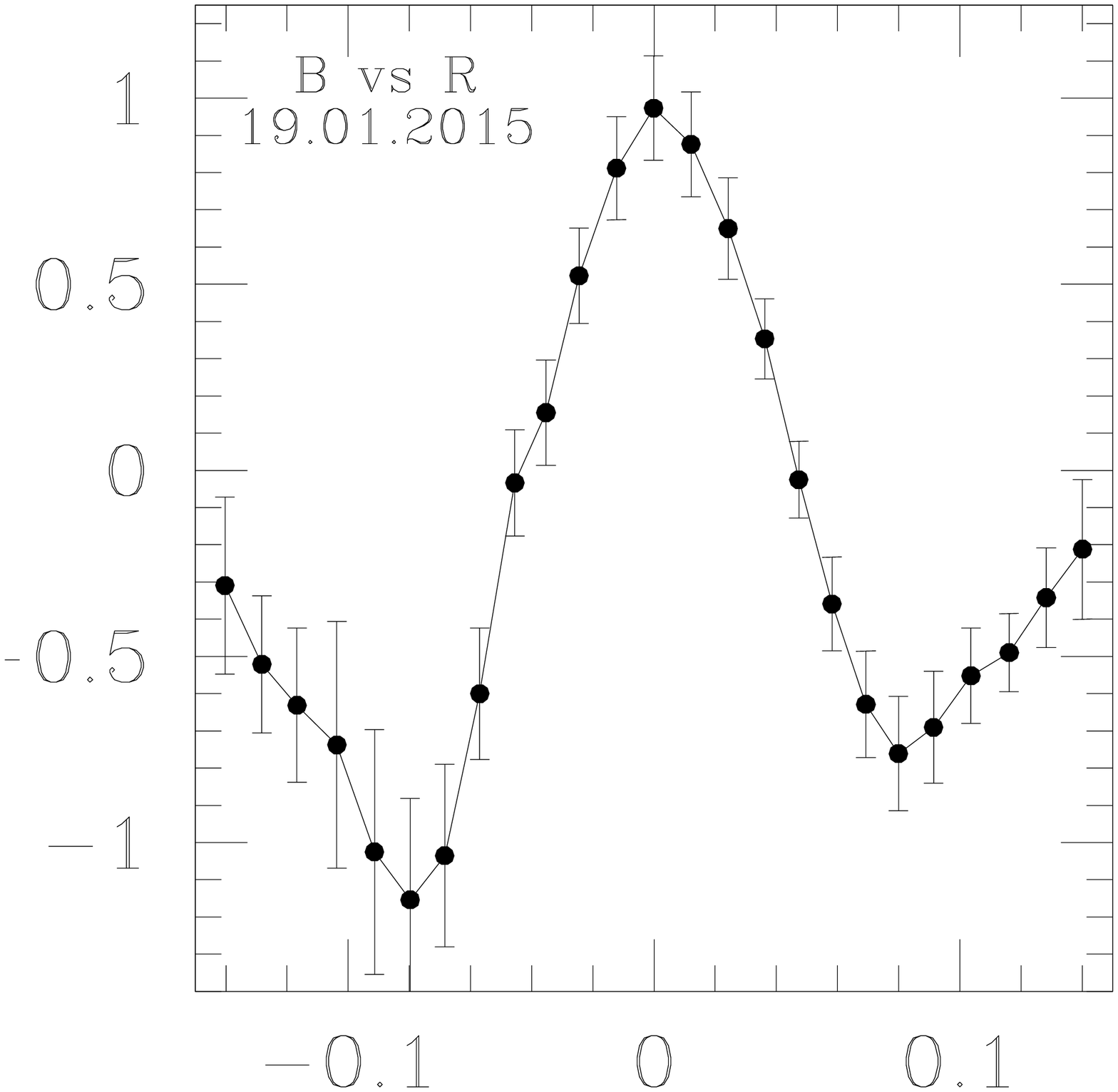,height=1.567in,width=1.59in,angle=0}
\epsfig{figure=  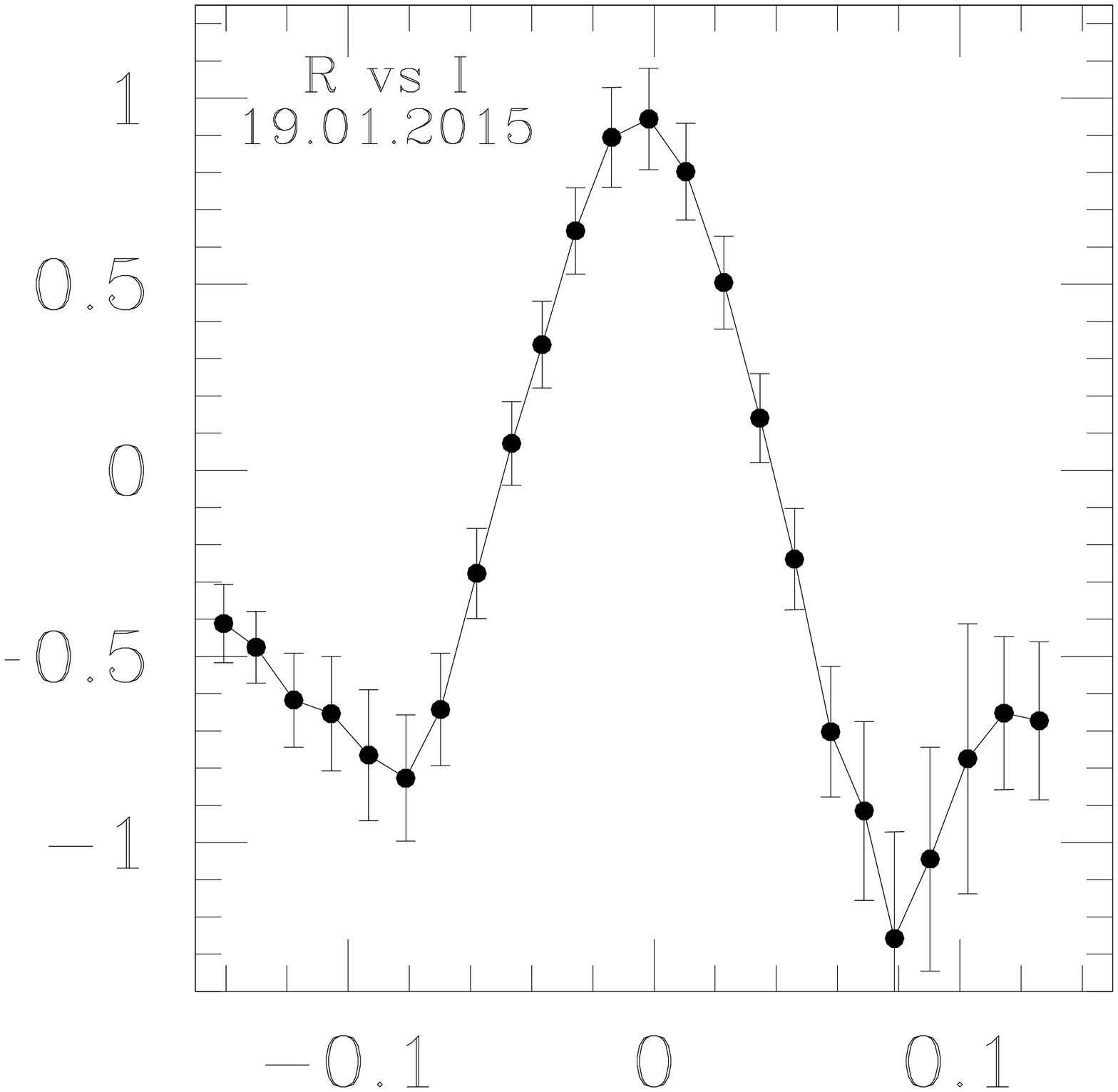,height=1.567in,width=1.59in,angle=0}
\epsfig{figure=  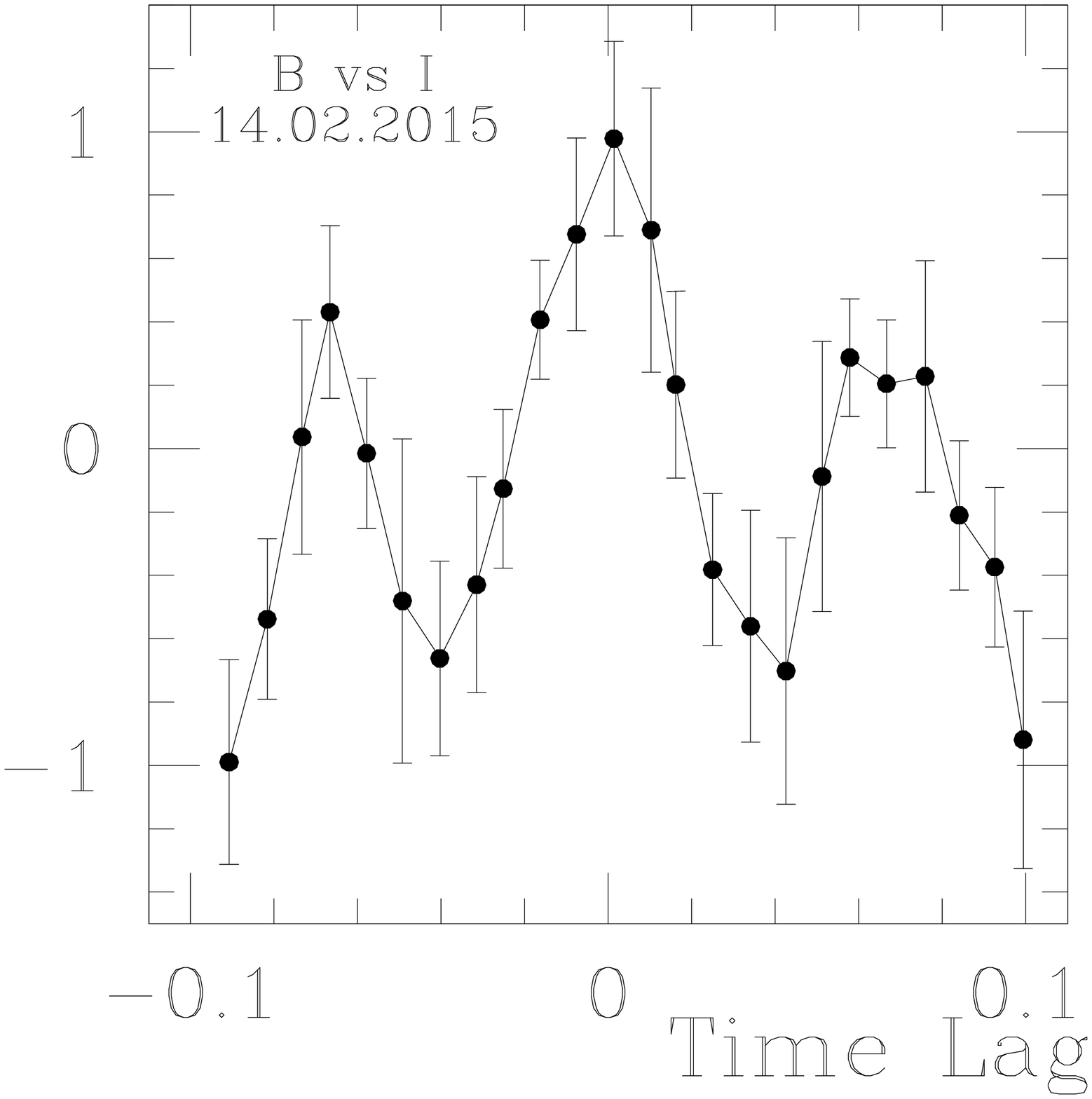,height=1.567in,width=1.59in,angle=0}
 \epsfig{figure=  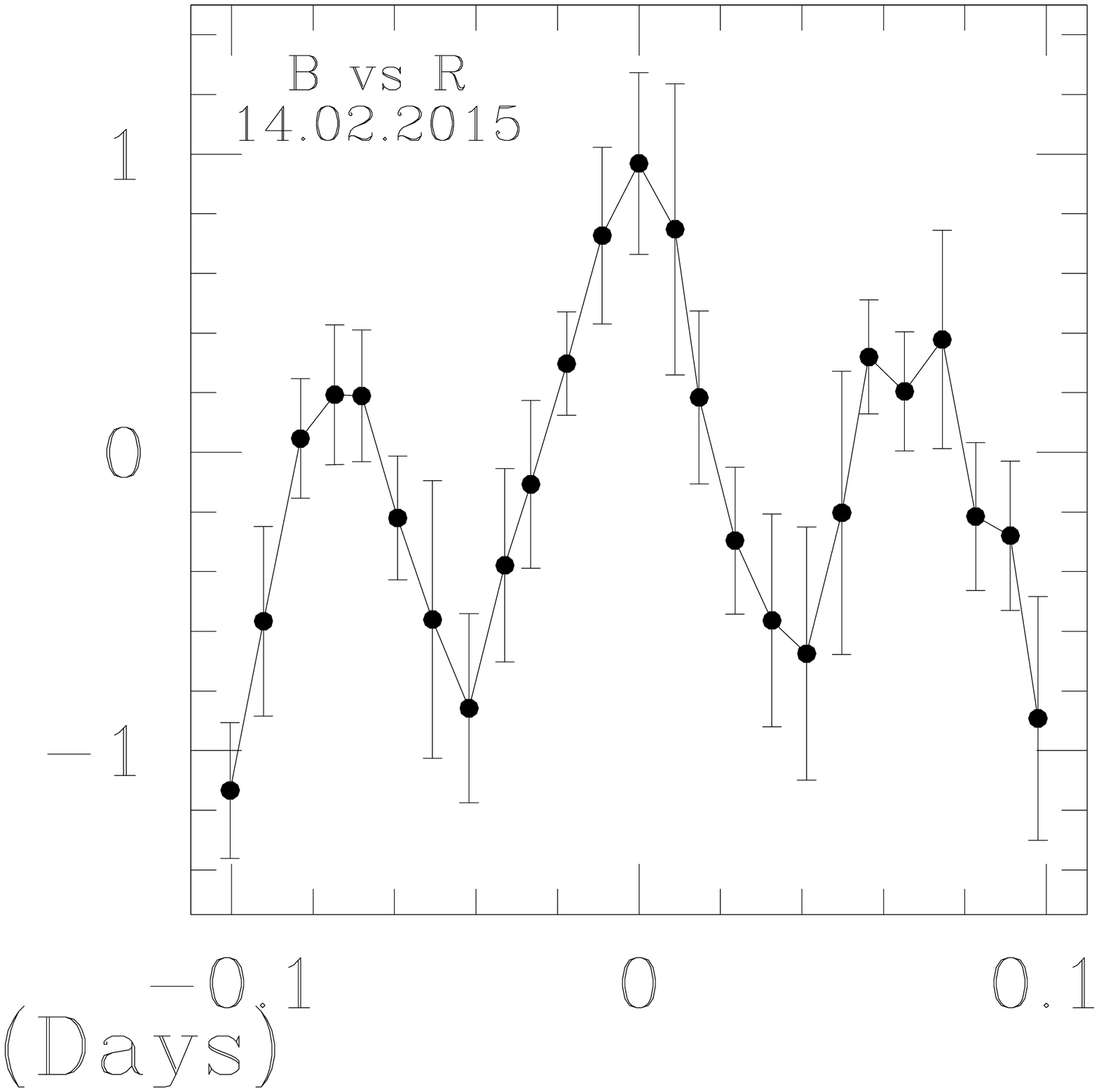,height=1.567in,width=1.59in,angle=0}
\epsfig{figure=  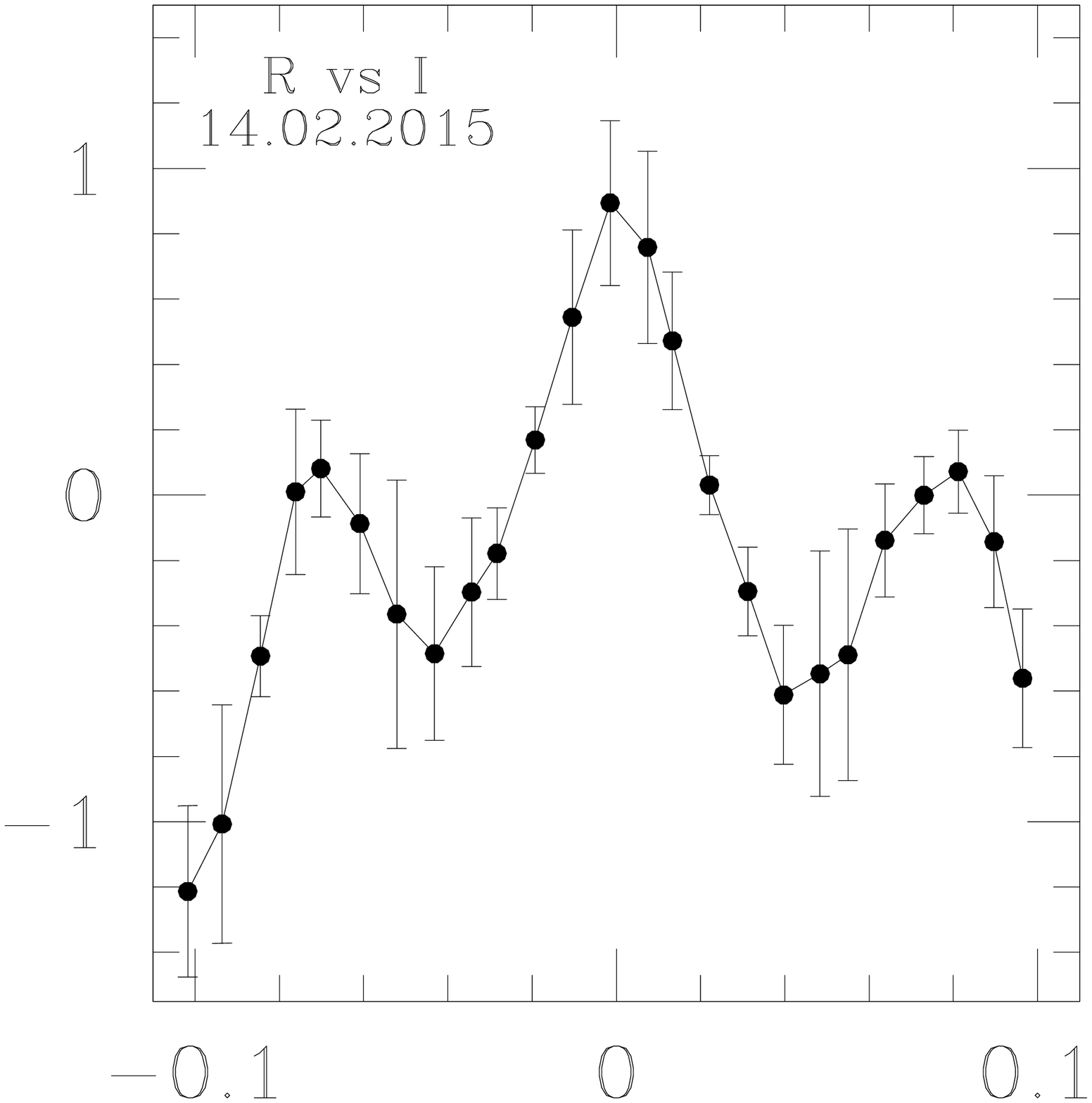,height=1.567in,width=1.59in,angle=0}

  \caption{DCF plots for S5 0716+714. In each plot, X and Y axis are the Time lags and DCF values, respectively.
Observation date is indicated in each plot.}
\end{figure*}

\subsection{Duty cycle}

The duty cycle (DC) for S5 0716+714 was calculated following the definition of Romero, Cellone, 
\& Combi (1999) that was later used by several authors (e.g.,  Stalin et al.\ 2009).
While calculating DCs we considered LCs with monitoring duration of at least two hours and by following the equation:

\begin{equation} 
DC = 100\frac{\sum_\mathbf{i=1}^\mathbf{n} N_i(1/\Delta t_i)}{\sum_\mathbf{i=1}^\mathbf{n}(1/\Delta t_i)} {\rm per cent} ,
\label{eq:dc} 
\end{equation} 
where $\Delta t_i = \Delta t_{i,obs}(1+z)^{-1}$ is the duration of the
monitoring session of the source on the $i^{th}$ night, corrected for
its cosmological redshift, $z$. Since the observing run time for the source is not the same for different night's observations, thus
the computation of the DC has been weighted by the actual monitoring duration
$\Delta t_i$ on the $i^{th}$ night. $N_i$ was set equal to 1  if
IDV was detected, otherwise $N_i$ = 0. We found the DC of our source to be $\sim$ 90\% when considering the LCs displaying
clear signature of variability (including the one PV case).
Nesci, Massaro \& Montagni (2002) found 80\% detectable variations in 52 observation nights.
All these observations indicate the source to have comparably high DCs.

\subsection{\bf Intra- and Inter-band correlations}

We computed DCFs to determine correlations between the B and I, V and I, and R and I bands and found close correlations among them with negligible
time lags in all nights where genuine variability is present. Unsurprisingly, for the nights when no genuine variability was present we found weaker or no
correlation among B, V, R, and I bands. We have not found any apparent timelag among the optical bands for this target. DCF plots are displayed
in figure 4. Wu et al.\ (2005) also
studied the correlated variability properties of this source and found no time lag among V and R variations.
The absence of time lags obtained during our observations in optical bands is likely due to the closeness of the bands, but does indicate
that photons in these frequencies are emitted from the same spatial region and are produced the by same physical process.
Following Section 3.2, we performed a SF analysis to see if there was
a discernable time scale of variability. The plots of SF analyses are presented in Figure 3.
The SF plot for 2014 Nov 15 starts with a continuous rising trend reaching a maximum, followed by a dip indicating the presence of a
nominal timescale of variability to be 273.6 minutes. While for the LC of Nov 21, the SF plot also increases, reaching a maximum, followed with a dip; the SF then
rises again, thus giving a possible variability timescales of 187.2 minutes. The first dips might provide an evidence of periodicity, but
as subsequent dips are not present, we cannot claim the significance of timescales or periodicities obtained using SF.

Since BL Lacertae objects have featureless optical continuum, their central
BH mass estimation through optical spectroscopy is not possible and so other approaches are necessary.
When no strong periodicity is detected we can  use these nominal intra-day timescales for this purpose if we assume
they correspond to  fluctuations in the inner portions of  the accretion disk (AD).
Explicitly, if one takes them to arise at a distance of $R = 5R_{s}$, where   $R_{s} = 2GM_{BH}/c^2$ is the Schwarzschild radius,
then the mass of SMBH can be estimated by (e.g.\ Gupta et al.\ 2012):

\begin{figure*}
\epsfig{figure=  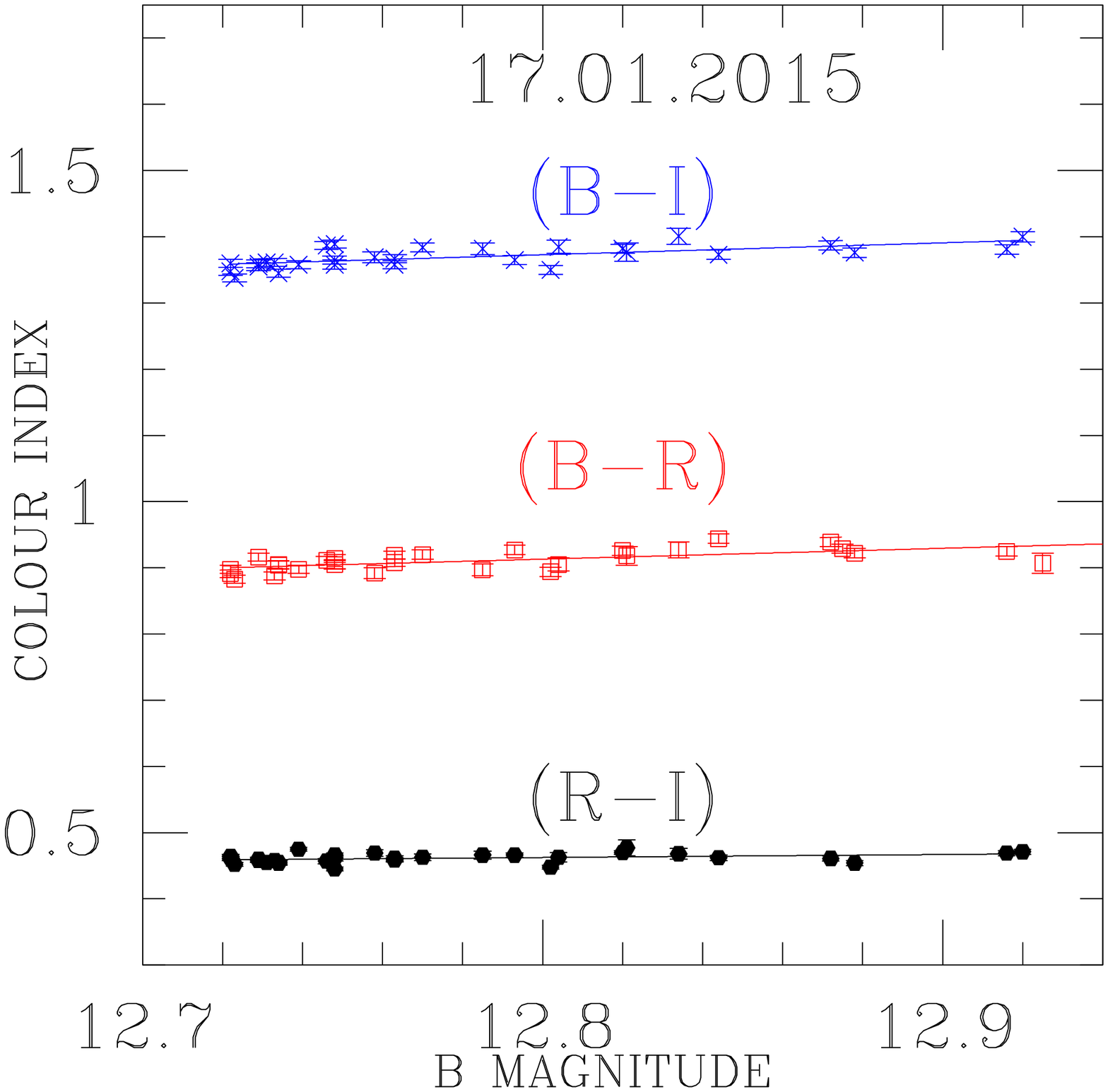,height=2.00in,width=2.26in,angle=0}
 \epsfig{figure=  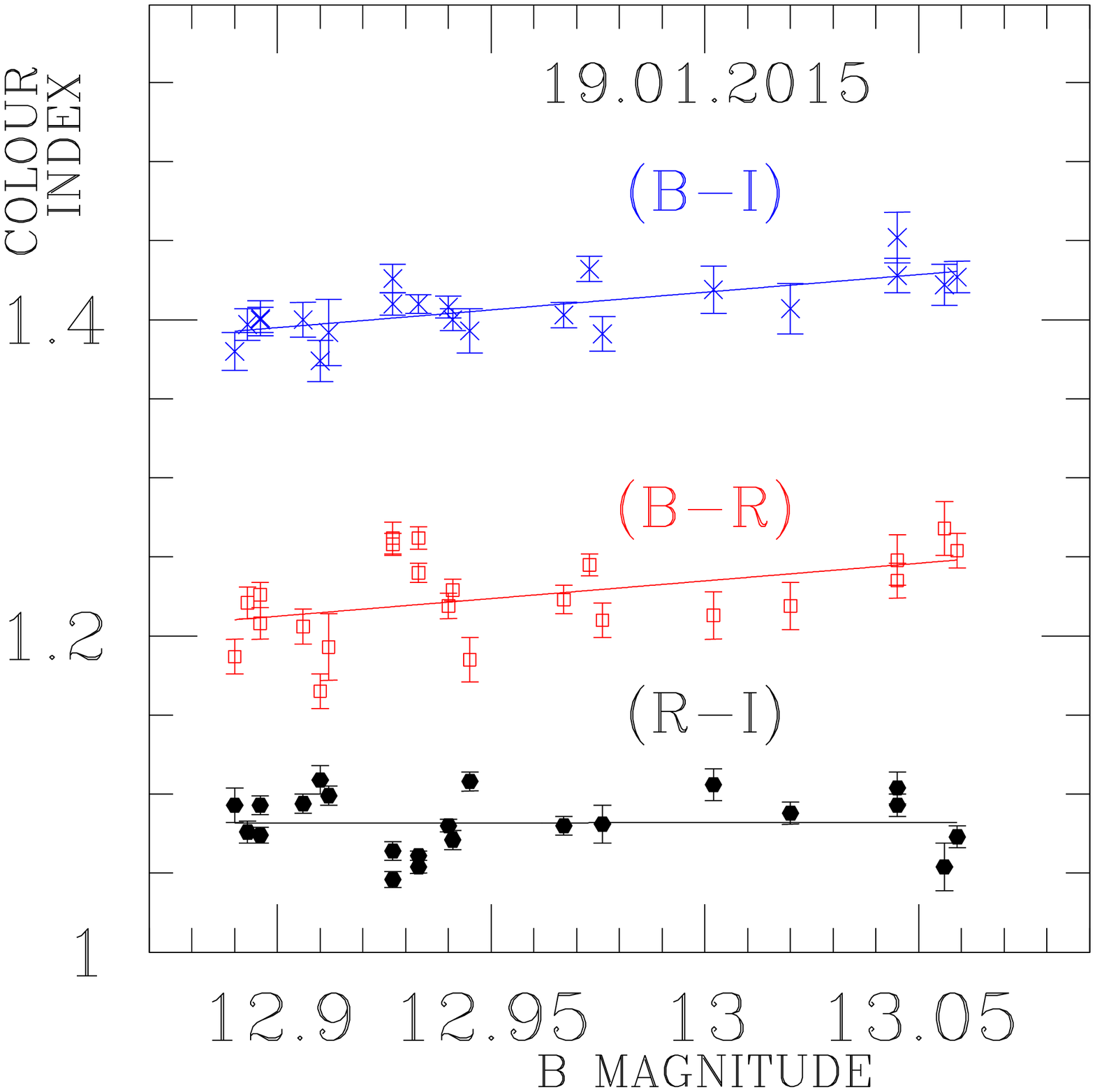,height=2.00in,width=2.26in,angle=0}
\epsfig{figure= 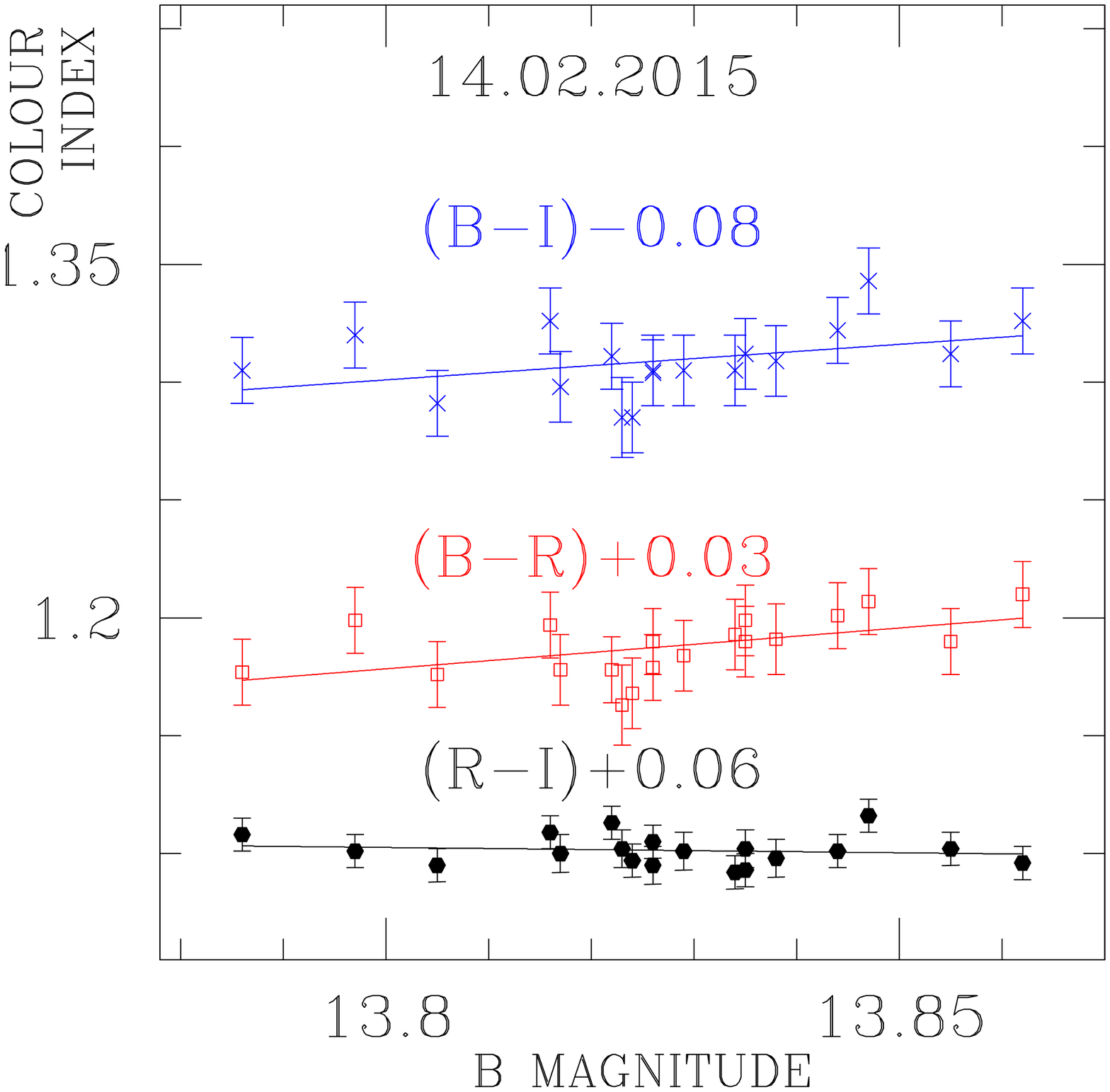 ,height=2.00in,width=2.26in,angle=0}
  \caption{ Color-index against magnitude on intraday timescales for S5 0716+714.}
\end{figure*}

\begin{equation}
M_{BH}= \frac{c^{3} \Delta t}{10 G(1+z)}
\end{equation}
If, as is far more likely for a blazar, the flux variations arise from the relativistic Doppler boosted jets, but we also assume that
these fluctuations are engendered by the AD and then advected into the jets,  we must multiply the M$_{BH}$ obtained above by an additional
Doppler boosting factor $\delta$ for an alternative estimate: $M_{BH}(\delta) = \delta  M_{BH} $ (e.g.\ Dai et al.\ 2007).
Of course, these mass estimates are invalid if the variations arise in the jets and are not explicitly related to the inner region
of the AD.  In that case any variability timescales provide constraints on the sizes of the emitting regions within
the jet.

Here we have used optical IDV timescales to derive  sizes of the emitting regions. 
For our target, as mentioned above, we found possible variability timescales of 273.6 min
and 187.2 min on 2014 Nov 15 and 21, respectively.  For them, $M_{BH}$ is calculated to be $2.55 \times 10^8$M$_{\odot}$
and $1.74 \times 10^8$M$_{\odot}$;  when the Doppler factor is taken into consideration, we get M$_{BH}(\delta)  = 2.42 \times 10^9$M$_{\odot}$
and $1.65 \times 10^9$M$_{\odot}$, where we have used $\delta$ = 9.49 (Zeng \& Zhang, 2011).

\begin{table*}
\caption{Fits to colour-magnitude dependencies and colour-magnitude correlation coefficients}

\noindent
\begin{tabular}{p{1.7cm}p{1.85cm}p{1.85cm}p{1.85cm}p{2cm}p{1.8cm}p{1.85cm}} \hline
Date of   &\multicolumn {2} {c} {B$-$R vs B}    &\multicolumn {2} {c} {R$-$I vs B} &\multicolumn {2} {c} {B$-$I vs B} \\
observation               & ~~~~~     $m^a$      & ~~~~~ $c^a$                   & ~~~~~ $m $       &  ~~~~~ $c $           & ~~~~~  $m$      &  ~~~~~  $c$           \\
      &  ~~~~~  $r^a$      & ~~~~~   $p^a$            & ~~~~~  $r$        & ~~~~~  $p$               & ~~~~~  $r$        & ~~~~~  $p$               \\\hline
17.01.2015 & 0.164 $\pm$ 0.005 & $-$1.190 $\pm$ 0.067 & ~~~0.045 $\pm$ 0.003   & $-$0.111 $\pm$ 0.045 & 0.181 $\pm$ 0.006 & $-$0.939 $\pm$ 0.081  \\
      & 0.676 $\pm$ 0.024 & $<$0.0001  & ~~~0.351 $\pm$ 0.027 & ~~~0.057  & 0.654 $\pm$ 0.026 & ~~~0.0001   \\
19.01.2015 & 0.223 $\pm$ 0.017 & $-$1.670 $\pm$ 0.221 & ~~~0.002 $\pm$ 0.013 & ~~~1.055 $\pm$ 0.173  & 0.224 $\pm$ 0.009 & $-$1.490 $\pm$ 0.245  \\
            & 0.456 $\pm$ 0.042 & ~~~0.029 & ~~~0.006 $\pm$ 0.045 & ~~~0.9778    & 0.698 $\pm$ 0.033 & ~~~0.0003    \\
 14.02.2015 & 0.346 $\pm$ 0.029 & $-$3.603 $\pm$ 0.395 & $-$0.045 $\pm$ 0.017 & ~~~1.723 $\pm$ 0.164   & 0.301 $\pm$ 0.034 & $-$2.860 $\pm$ 0.183    \\
 	     &0.506 $\pm$ 0.055 & ~~~0.027 & $-$0.129 $\pm$ 0.048 & ~~~0.598 & 0.392 $\pm$ 0.037  & ~~~0.097 \\
\hline
\end{tabular}  \\
$^a$ $m =$ slope and $c =$ intercept of CI against B; $r =$ Pearson coefficient; $p =$ null hypothesis probability\\
\end{table*}

\subsection{\bf Color-magnitude relationship}

Optical flux variations are accompanied by spectral changes which can be further analyzed by finding colour indices.
We now look for  any relationship between the colour indices of the source and the brightness in the B band.

Color--magnitude plots for three nights, 2015 Jan 17, Jan 19 and Feb 14 are displayed in Figure 5. For each night we fitted the plots
of each colour index, CI, against B magnitude with straight lines, i.e., CI =$m$B + $c$. The best fit values of $m$ and
$c$, are listed in Table 4 along with the linear Pearson correlation coefficient, $r$, and the corresponding null 
hypothesis probability, $p$. A positive slope here implies significant positive correlation between CI and apparent blazar magnitude
when Pearson's correlation coefficient, $r$ $\sim 0.9$; indicating that the source tends to be bluer when it brightens (BWB) or redder when it dims. 
We did not find any significant correlation (Pearson's correlation coefficient, $r$ $\sim 0.9$)
between B band magnitude and colour indices on both intraday and several month timescales.
The latter shown in Figure 6 and the respective best fit values listed in Table 5. 

Similar to our colour magnitude analysis, Raiteri et al.\ (2003) also found at most weak colour-magnitude correlation i.e. spectrum steepening
with source brightness decreasing was detected on few instances only. Also, Wu et al.\ (2005) found no spectral changes with magnitude
variations during their one night monitoring.
Our results are consistent with Stalin et al.\ (2006). They also found no evidence of spectral changes with the source brightness on either
internight or intranight timescales for the BL Lacertae object S5 0716+714 even when the target was in flaring state.
No clear colour trends have been reported by many authors in several cases (e.g. Ghosh et al.
2000; B{\"o}ttcher et al. 2007; Poon et al. 2009).

\begin{figure}
\centering
\epsfig{figure=  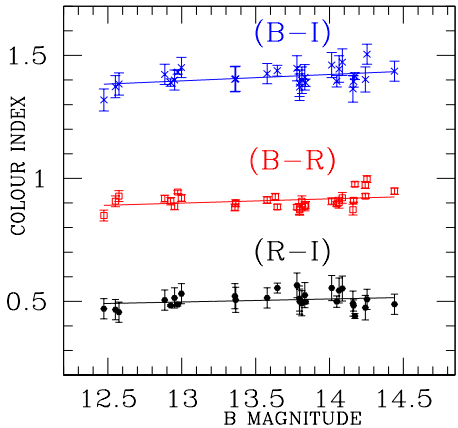,height=2.263in,width=2.401in,angle=0}
  \caption{ Color--magnitude plots on short timescales for S5 0716+714.}
\end{figure}

\begin{table}
\caption{Color-magnitude dependencies and colour-magnitude correlation coefficients on short timescales.  }
\textwidth=7.0in
\textheight=10.0in
\vspace*{0.2in}
\noindent

\begin{tabular}{ccccc} \hline 

Color Indices     &  $m_1^a$  &  $c_1^a$  &   $r_1^a$  & $p_1^a$    \\ \hline
(B-R)          &  0.017 & 0.674 & 0.304  & 0.102  \\     
(R-I)          &  0.012 & 0.339 & 0.230 & 0.238  \\ 
(B-I)          &  0.026 & 1.060 & 0.403 & 0.033 \\
\hline
\end{tabular}\\
$^a$ $m_1 =$ slope and $c_1 =$ intercept of CI against V; \\
$r_1 =$ Pearson coefficient; $p_1 =$ null hypothesis probability \\
\end{table}

Resolving the colour trend in blazars on intranight timescales can help us investigate the origin of blazar emissions and also constrain
various blazar variability models.
Any AD radiation is expected to be overwhelmed by that from the strongly Doppler boosted jets, thus the observed colour variations cannot be
explained by AD based models. Also, the underlying host galaxy of the target is more than 4 times fainter than the brightness of the
target itself (Nilson et al.\ 2008), thus colour variations due to variable contributions from the host galaxy (due to changes in seeing) are negligible (Hawkins 2002).
According to shock-in-jet models, shocks propagating down the Doppler boosted relativistic jets accelerate particles and
compress magnetic fields thus causing the flux and spectral variations (e.g.\ Marscher \& Gear 1985; Marscher et al.\ 2008). The shock thickness determines
variability amplitude and timescales.
During our observations, the amplitude of variations is systematically larger at higher frequencies.
Color--magnitude variations can be explained by both shock-in-jet models and geometric effects.
Sun et al.\ (2014) found that colour variability in quasars is timescale dependent. Their timescale dependent model also militates against the possibility of the
BWB trend being due to mixture of a variable disk emission with a blue but constant colour, and a redder stable emission, as from the host galaxy.

The flux spectral index also can be used to analyze optical flux variations in blazars as they are usually associated with the changes in the spectral
shape. The average spectral indices are calculated simply as (e.g., Wierzcholska et al.\ 2015),
\begin{equation}
\langle \alpha_{BR} \rangle = {0.4\, \langle B-R \rangle \over \log(\nu_B / \nu_R)} \, ,
\end{equation}
where $\nu_B$ and $\nu_R$  are effective frequencies of the respective bands (Bessell, Castelli, 
\& Plez 1998).
The optical slope for three nights, when quasi-simultaneous observations in B and R filters were taken,
was calculated as 2.21 $\pm$ 0.02 (Jan 17), 2.24 $\pm$ 0.03 (Jan 19), and 2.15 $\pm$ 0.03 (Feb 14).
For the other  20 nights, the optical slope varied only slightly between $\alpha_{BR}$ $\sim$ 2.1 to 2.4 as shown in Figure 7.
These relatively steep spectral indices  indicate strong
synchrotron emission from the blazar jet and a very small AD contribution.

\begin{figure}
\epsfig{figure= 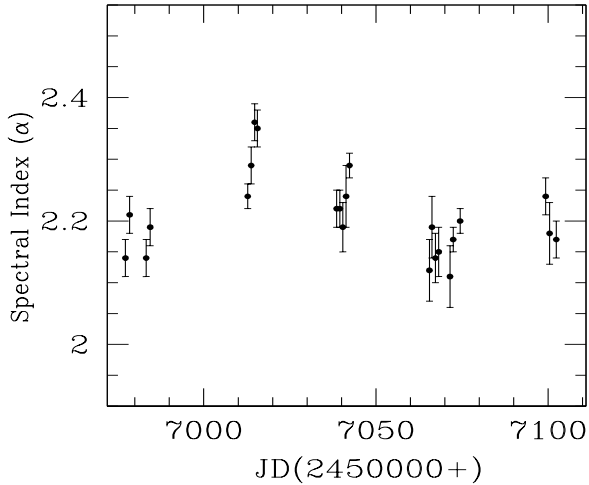,height=2.6in,width=3.1in,angle=0}
\caption{Variation of average optical spectral index calculated using equation 6 Vs time covering the entire observation period for the target.}
\end{figure}

\subsection{Spectral Energy Distribution (SED)}

To learn more about the emitting region and the processes that energize the radiating electrons, one can generate SEDs at different stages
of the outburst as they can be crucial in analyzing individual emission components.   Modeling of SEDs at different times  is an excellent diagnostic tool to distinguish
between the nonthermal emission from the relativistic jets or the thermal emission from the AD, or broad line region (BLR), or
hot dust at parsec scales.  Broadband SEDs can further be used to derive information on the physical parameters of the emitting region
such as  magnetic field, electron energy density, and
changes in the Doppler factor.  While we are limited in constructing SEDs, as we only have fluxes in the optical band, it is nonetheless
worthwhile to examine them.

In this analysis, we de-reddened the calibrated magnitudes of the BL Lac S5 0716+714 by subtracting Galactic
absorption using the NED extinction calculator\footnote{\tt http://ned.ipac.caltech.edu}
with the following values: $A_{B}$ = 0.112 mag, $A_{V}$ = 0.085 mag, $A_{R}$ = 0.067 mag,
and $A_{I}$ =  0.047 mag (Cardelli et al.\ 1989; Bessell, Castelli, \& Plez 1998).
Quasi-simultaneous narrow band SEDs generated using B, V, R, and I data sets for our source corresponding to 23 epochs between 2014 - 2015 are
displayed in Fig.\ 8. The faintest fluxes for S5 0716+714 were measured on 2014 Dec 22 while they reached a maximum on 2015 Jan 18;
strong variations are seen in the SEDs
during our observational span between 2014 Nov -- 2015 March.

\begin{figure}
\centering
\epsfig{figure= 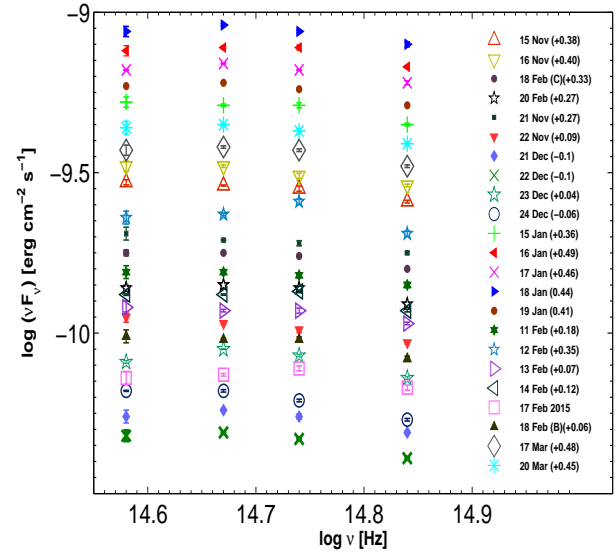,height=3.1in,width=3.5in,angle=0}
\caption{SED results for S5 0716+714 in optical frequency range.}
\end{figure}

During the GASP-WEBT-AGILE campaign of 2007, Villata et al.\ (2008) studied the optical-IR SED of  S5 0716$+$714. Modeling of the SED
of this blazar  was carried out by Giommi et al.\ (2008) using two SSC emission models, representing a slowly
and a rapidly variable component, respectively.
Understanding extreme variability of the blazars at all wavelengths can provide information about the processes operating in the inner regions
of the source.  However, quasi-simultaneous observations at X-ray, gamma-ray, cm and mm wavelengths along with the optical are most helpful in
conducting more
detailed studies of the spectral evolution of the target
during this flaring period.

\section{Discussion and Conclusion}

Since the synchrotron peak of this blazar's SED is located close to the optical wavelengths, studies in these bands can yield information about
the presence of components other than synchrotron continuum which could include thermal emission from the AD, radiation from the
region surrounding the nucleus and host galaxy contamination. Blazar variability studies have been very useful in understanding their nature
and extreme conditions within the emission region.
Blazar emission from radio to optical is predominantly synchrotron radiation from the ultra-relativistic electrons gyrating in the magnetic field
of the Doppler boosted relativistic jets (e.g.\ Marscher \& Gear 1985). Doppler boosting of the non thermal emission from the relativistic jet shortens
the observed variability timescales and increases the observed variability amplitude.

Intrinsic origins for AGN variability involves two major classes of models, i.e., shock-in-jet based models (e.g.\ Marscher \& Gear 1985) and
the AD based models (e.g.\ Chakrabarti \& Wiita 1993).
Jet based models of RLAGNs certainly are favored for the production of  blazar variability during their outburst states as then the blazar emission is mainly
nonthermal Doppler boosted relativistic jet emissions (Blandford \& Rees 1978; Hughes et al.\ 1992).  These variations result from developing/decaying
shocks in the jets (e.g.\ Marscher \& Gear 1985; Hughes, Aller, \& Aller 2011) or from helical motions 
(e.g.\ Camenzind \& Krockenberger 1992) or from turbulence behind these shocks
(e.g.\ Marscher 2014; Calafut \& Wiita 2015) or from changing jet directions (e.g.\ Gopal-Krishna \& Wiita 1992).
A relativistic shock propagating down a jet will accelerate electrons even to TeV energies where the shock interacts with a region of high
magnetic field or electron density, thus causing radiations at different frequencies being produced at different distances. 
Frequency dependence of the duration of the flare then corresponds to an energy dependent cooling length behind the shock front thus causing colour variations in blazars.
The observations during early rising phase of the flux will give a bluer colour while those taken during later phases of the same flare will show
more enhanced redder fluxes (Papadakis et al.\ 2007). 

Presence of BWB/RWB trends in blazars can be explained due to superposition of both blue and red emission components where the red contribution can be
attributed to the synchrotron radiation from the jet while the blue one could come from the thermal emission from the AD around the central region.
 The presence (or absence) of correlation between colour indices and magnitude at different timescales can help us probe physical processes
 responsible for blazar variability. During quasi-simultaneous multi-wavelength obserations, possibility arises that source brightness might
 change when switching observations from one band to another. Thus to explore the spectral changes in the target, very dense
 simultaneous multi-band monitoring with high precision will
 prove to be fruitful.

Variability studies on IDV timescales can give an upper limit to the size of the emitting region and thence perhaps give an estimate of the SMBH mass
residing at the centre of the galaxy (e.g. Gupta et al.\ 2008) which could in turn shed some light on the evolution process in AGNs
(Barth 2002; Fan 2003).
The shortest timescale of variability is associated with the light crossing time of the variable region, which in AD based models is directly proportional to the BH masses
(Bachev, Strigachev, \& Semkov 2005).
The probability of detecting variability in blazars increases with the duration of observations.  For example,
64\% of blazars were found to show IDV when the duration of observations was $\leq$ 3 hr but this increased to 82\% when the observations lasted
for $>$ 6 hr (Gupta \& Joshi 2005). Duty cycle of variations in blazars is $>$ 92\% on short term basis while it reaches $\sim$ 100\% on long
term basis confirming that variability detection increases with the duration of observations (Rani et al.\ 2010).

In this paper, we reported the results of quasi-simultaneous optical observations of the BL Lacertae object
S5 0716+714 in the B, V, R and I filters during 2014 Nov -- 2015 March using 2 telescopes in Bulgaria and 1 in India.
We measured multiband optical flux, colour and spectral variations in the target on intraday and short term timescales.
During the analyzed period, the source attained the brightest magnitude ever recorded. If we include the PV case, 
then we can say that confirmed IDV was detected on all nights with a maximum
variability amplitude of 22.28\% in B passband on 2015 Jan 01. Amplitude of variability was found to increase with frequency, as was also found by
Papadakis et al.\ (2003). Using DCF analysis we found that the LCs in B, V, R and I are well correlated in most of the cases with no significant
temporal lags that indicate a the  distance between emitting regions (e.g.\ Raiteri et al.\ 2003). Since we are getting null time lags in optical bands,
 we can say that the optical emissions of different wavelengths arise from the same region.

A SF analysis gave nominal variability timescales of 273.6 min
and 187.2 min  on 15 and 21 Nov 2014, respectively, for which a $M_{BH}$ would be estimated to be $2.55 \times 10^8$M$_{\odot}$
and $1.74 \times 10^8$M$_{\odot}$ in the simplest, albeit, unlikely interpretation that they arise from the AD.
 We have studied the corresponding variations in the (B-I), (B-R), and (R-I) colour indices of the source with respect
to brightness variations in B band. Our observations did not
reveal the presence of spectral variability in the target during its high state on intraday or short timescales.
The source was highly variable with a duty cycle of $\sim$ 90\% when all filters were considered.
Variability amplitude was greater in higher frequency bands, which is consistent with previous studies (Papadakis et al.\ 2003; Hu et al.\ 2006).
The optical spectral index of the source varied only slightly between $\alpha$ $\sim$ 2.1 and 2.4, thus revealing a steep optical spectrum. 
We constructed a quasi-simultaneous SEDs of our source at different times and saw the peak frequency shift.
In case of BL Lacertae objects, the optical band is near the peak of the synchrotron component of the SED, so optical variability directly reflects
acceleration/cooling processes acting on the highest energy electrons.   We still do not have sufficient information about basic blazar parameters such as jet composition, shock formation, and beaming parameters 
that  could further help in understanding the emission regions of the blazars. Therefore variability studies of a large sample of blazars on diverse
timescales remain critically important.


\section*{Acknowledgments}
We thank the referee for detailed and thoughtful comments which helped us to improve the manuscript. 
JHF thanks the support on grant NSFC No  U1531245.
This research was partially supported by Scientific Research Fund of the Bulgarian Ministry of
Education and Sciences under grant DO 02-137 (BIn-13/09).

\label{lastpage}
\end{document}